\documentclass[preprint,showkeys,secnumarabic,amsfonts,showpacs,amsmath,amssymb]{revtex4}
\usepackage{natbib}
\usepackage{bibentry}
\usepackage[dvips]{color}
\usepackage{epsfig}
\usepackage{amsmath}
\usepackage{graphicx}
\usepackage{amssymb,epsf}
\usepackage{epstopdf}
\usepackage{makeidx}
\usepackage{verbatim}
\usepackage{natbib}
\usepackage{hyperref}
\def\Box{\hbox{$\rlap{$\sqcup$}\sqcap$}}
\begin{document}
\title{ Stability of Modified Gravity Coupled by the Weyl Tensor}

\author{M. Ghanaatian}\email{m$_$ghanaatian@pnu.ac.ir}\affiliation{Department of Physics, Payame Noor University, Iran}
\author{A. Gharaati}\email{agharaati@pnu.ac.ir}\affiliation{Department of Physics, Payame Noor University, Iran}
\author{F. Milani}\email{f.m.1683@hotmail.com}\affiliation{Department of Physics, Payame Noor University, Iran}

\date{\today}

\begin{abstract}\label{sec:abstract}
\noindent \hspace{0.35cm}

In this paper, we try to consider the stability conditions of
a modified gravity coupled by Weyl tensor. In this way, we indicate the suitable conditions for a successful bounce
while the equation of state (EoS) parameter crosses the phantom divider for
our new corrected modified gravity. In the spatially flat Friedmann-Lema\^{\i}tre-Robertson-Walker  (FLRW)
universe, according to the ordinary version of the holographic dark
energy (HDE) model, describing accelerated expansion of the universe, will be
considered. Our model's stability conditions and its general properties of
attractors for scalar field scripts which control cosmological acceleration will be checked. The statefinder diagnostic parameters of our model will be compared by today's observational data and defined methods.    
However its EoS parameter will be obtained, too.
\end{abstract}



\maketitle

\section{Introduction}\label{sec:Introduction}
Since ancient centuries, philosophers, scientist and specially cosmologists try to explain how our universe has
formed? Albert Einstein's
theory of general relativity, modified gravity, big bang and big bounce theories are alternative
ways of looking at how the universe has commenced. In physical cosmology,
the age of the universe is defined as the elapsed time since the big bang and it
tries to enunciate what the universe looked like before the
planets and the stars came to existence.

By current project of the European space agency (ESA) which called Planck \cite{planck2015planck}, the age of the universe is measured  $13.799\pm0.021$ billion years
within the Lambda cold dark matter ($\Lambda$CDM) model which describes the evolution of the universe, has extended from a very uniform, hot, dense primordial state to its present state.
Since 1965, the cosmic microwave background (CMB) has played a
central role in cosmology. The discovery of the CMB confirmed
a major prediction of the big bang theory and was difficult to reconcile with the steady
state theory \cite{penzias1965measurement}. The precision measurement of the CMB spectrum by NASA's cosmic background
explorer (COBE) \cite{mather1990preliminary,mather1994measurement} mission confirmed the predicted CMB blackbody spectrum, which results from thermal equilibrium between matter and radiation
in the hot, dense early universe. The COBE detection of CMB anisotropy \cite{smoot1992structure}\nocite{bennett1992preliminary,kogut1992cobe,wright1992interpretation} established the amplitude
of the primordial scalar fluctuations and supported the case for the gravitational evolution
of structure in the universe from primordial fluctuations.
In this way, $\Lambda$CDM model is well understood theoretically and strongly supported by recent high-precision astronomical observations such as  Wilkinson microwave anisotropy probe (WMAP) \cite{bennett2003first,bennett2011seven,bennett20141,bennett2013nine} and other probes.

Nowadays most of the theories and evidences tell us, the universe, in early time, was condensed
into a small region of matter and energy, which called a singularity. Moreover, cosmologists believe that the singularity suddenly, exploded and expanded at an
incredibly fast rate. The matter combined to create protogalaxies,
which, in turn, combined to form galaxies and it continued to form planets, etc. But why it was condensed? How is possible making all of things from nothing (no space, no time coordinates)? Some cosmologists try to find a suitable answer to these paradoxes. They suppose that the universe will eventually grow no
more when the first singularity is eliminated. So our universe will collapse in on itself into a quasi-singularity
as gravity pulls matter down, an event billions of years from now
which is called the big crunch. 
The cosmologists believe that unseen materials exist and may
exert enough gravitational force to stop the universe's expansion
and cause the big crunch. The bouncing theory combines the big
bang and big crunch theories to develop a vision of an infinite,
cyclical cosmos in which the universe over and again expands from
a quasi-singularity only to ultimately collapse back in on itself,
before doing it all over again. To put it another way, a bouncing
universe would continuously expand and contract {\cite{sadeghi2009bouncing,bamba2014bounce,farajollahi2010bouncing}}.

Undoubtedly observational data of
type-Ia Super-Novae (SNIa) {\cite{riess2004type,knop2003new}}, which is a type of them that occurs in binary systems -- two stars orbiting one another, in which one of the stars is a white dwarf -- have determined basic cosmological parameters in high-precisions. 
SNIa is among the most important probes of expansion and historically led to the general acceptance that a Dark Energy (DE) component is needed {\cite{riess1999bvri,perlmutter1999measurements,ade2015planck}. In physical cosmology and astronomy, there are indicative of the fact that, the universe is spatially flat because SNIa is a sub-category in the Minkowski-Zwicky supernova classification scheme and DE is an unknown form of energy which is hypothesized to permeate all of space, tending to accelerate the expansion of the universe. It is the most accepted hypothesis to explain the observations since the 1990s indicating that the universe is expanding at an accelerating rate. Assume the standard model of cosmology is correct, the best current measurements indicate that DE contributes $68.3\%$ of the total energy in the present-day observable universe. The mass–energy of Dark Matter (DM) and ordinary (baryonic) matter contribute $26.8\%$ and $4.9\%$, respectively, and other components such as neutrinos and photons contribute a very small amount \cite{ade2014planck,francis2013first}. Again on a mass–energy equivalence basis, the density of DE $(\sim 7 \times 10^{-30} g/cm^3)$ is very low, much less than the density of ordinary matter or DM within galaxies. However, it comes to dominate the mass–energy of the universe because it is uniform across space \cite{steinhardt2006cosmological,shrestha2015dark}.

Simultaneously, as to the origin of DE, they posed a
fundamental problem. The combined analysis of SNIa
{\cite{perlmutter1999measurements,riess1998observational,riess1999bvri}},
 contingent upon the background expansion
history of the universe around the redshift $z<1$ as galaxy
clusters measurements and Wilkinson Microwave Anisotropy Probe
(WMAP) data \cite{bennett2003first,bennett2011seven,bennett20141,bennett2013nine}, Sloan Digital Sky Survey (SDSS)
{\cite{abazajian2005third,abazajian2004second,abazajian2003first,tegmark2004three}}, Chandra X-ray Observatory (CXO)
{\cite{allen2004constraints}} etc. It shows some cross-checked information
of our universe, providing surprising proof as to the fact that
the expansion of the universe for the time being, seems to have
been accelerating behaviour, being imputed to DE with negative pressure. In contrast, DM, a matter without pressure, is basically utilized to describe
galactic curves and large-scale structure formation \cite{tonry2003cosmological,tegmark2004cosmological}.

It is shown by the cosmological acceleration, that the present day
universe is dominated by smoothly distributed slowly varying DE
component and it reinforces the big bouncing hypothesis. The constraint derived from SNIa has a degeneracy in
the equation of state (EoS) of DE {\cite{astier2006supernova,riess2007new,wood2007observational,bamba2012dark}}. However, the
nature of DE  is unknown until now but people have
suggested some candidates for its explanation. The cosmological
constant, $\Lambda$, in a model which the universe's equation
has a cosmological constant, indicated by $\Lambda$, and Cold Dark
Matter ($\Lambda$CDM), is the most notable theoretical candidate
of DE, which has an equation of state with $\omega = -1$. This
degeneracy is offered even by adding other constraints coming from
CMB {\cite{spergel2003first,spergel2007three}} and Baryon
Acoustic Oscillations (BAO) {\cite{eisenstein2005detection}}. Astronomical observations
denote that the cosmological constant, in their orders of
magnitude, tend to be much smaller than it is calculated in modern
theories of elementary particles {\cite{weinberg1989cosmological}}. Two of the most
notable difficulties faced with the cosmological constant are the
"fine-tuning" and the "cosmic coincidence" {\cite{copeland2006dynamics}}. The
constraints, nowadays, on the EoS around the cosmological constant
value, are $\omega= -1 \pm 0.1$ {\cite{tonry2003cosmological,tegmark2004cosmological,astier2006supernova,riess2007new,wood2007observational,
bamba2012dark,spergel2003first,spergel2007three,eisenstein2005detection,weinberg1989cosmological,
copeland2006dynamics,seljak2005cosmological,tegmark2005does}} and
this probability exists that $\omega$ may differ in time {\cite{sadeghi2008non,sadeghi2009crossing,sadeghi2009bouncing,farajollahi2010bouncing,farajollahi2011cosmic,farajollahi2011stability,
farajollahi2012cosmological}}. From the
theoretical point of view there are three essentially different
cases: $\omega>-1$ (quintessence), $\omega = -1$ (cosmological
constant) and $\omega <-1$ (phantom)
({\cite{caldwell2002phantom,caldwell2003phantom, carroll2003can, cline2004phantom, mcinnes2002ds, melchiorri2003state, nesseris2004fate, onemli2004quantum, alam2004case, padmanabhan2002accelerated, padmanabhan2002can, hao2003attractor, singh2003cosmological, carroll2005can, aref2006exact,guo2005interacting}} and refs. therein).

The models of DE can be generally categorized into two groups which their classifications have been considered in our previous paper \cite{ghanaatian2014bouncing}. So in this paper we try to use the modified gravity by replacing some additional terms in the Einstein-Hilbert action \cite{Feynman1995Feynman} which is the action that yields the Einstein field equations through the principle of least action. With the $(- \, + \, + \, +)$ metric signature, the gravitational part of the action is given as $I_{EH}=\frac{1}{16\pi G}\int{d^4 x\sqrt{-g}R}$ where $G$ is the gravitational constant. 
Then we use the Friedman equation to constitute the starting
point as all researches in cosmology. However the Friedman
equations have been corrected during the past few years being
proposed in varying contexts, generally inspired by brane-world
investigation {\cite{copeland2005generalized, tsujikawa2004unified}}. These changes are often of a type
that involves the total energy density $\rho$. In {\cite{sadeghi2008non,townsend2004cosmology}},
multi-scalar coupled to gravity is studied in the context of
conventional Friedman cosmology. It is found that the cosmological
trajectories can be viewed as geodesic motion in an increased
target space.

In this ways, there are several phenomenological models which describe the
crossing of the cosmological constant barrier {\cite{boisseau2000reconstruction,sahni2003braneworld,mcinnes2005phantom,
perivolaropoulos2005constraints,feng2005dark, guo2005cosmological,anisimov2005b,witten1986non,witten1986interacting}}. Therefore finding a model following from the basic principles
describing a crossing of the $\omega = -1$ barrier will be one of our 
important and essential goals. 

Moreover, in mathematics, stability theory addresses the stability of solutions of differential equations and of trajectories of dynamical systems under small perturbations of initial conditions. In dynamical systems, an orbit is called Lyapunov stable if the forward orbit of any point is in a small enough neighbourhood or it stays in a small neighbourhood. Various criteria have been developed to prove stability or instability of an orbit. Under favourable circumstances, the question may be reduced to a well-studied problem involving eigenvalues of matrices. Many parts of the qualitative theory of differential equations and dynamical systems deal with asymptotic properties of solutions and the trajectories, what happens with the system after a long period of time. The simplest kind of behaviour is exhibited by equilibrium points, or fixed points, and by periodic orbits. If a particular orbit is well understood, it is natural to ask next whether a small change in the initial condition will lead to similar behaviour. Stability theory addresses the following questions: Will a nearby orbit indefinitely stay close to a given orbit? Will it converge to the given orbit? In the former case, the orbit is called stable and in the latter case, it is called asymptotically stable and the given orbit is said to be attracting.

Stability means that the trajectories do not change too much under small perturbations. In general, perturbing the initial state in some directions results in the trajectory asymptotically approaching the given one and in other directions to the trajectory getting away from it. There may also be directions for which the behaviour of the perturbed orbit is more complicated, and then stability theory does not give sufficient information about the dynamics.

One of the key ideas in stability theory is that the qualitative behaviour of an orbit under perturbations can be analysed using the linearization of the system near the orbit. In particular, at each equilibrium of a smooth dynamical system with an $n$-dimensional phase space, there is a certain $n\times n$ matrix $A$ whose eigenvalues characterize the behaviour of the nearby points \cite{hartman1960lemma,hartman1960local,grobman1959homeomorphism}. More precisely, if all eigenvalues are negative real numbers or complex numbers with negative real parts then the point is a stable attracting fixed point, and the nearby points converge to it at an exponential rate. If none of the eigenvalues are purely imaginary (or zero) then the attracting and repelling directions are related to the eigenspaces of the matrix $A$ with eigenvalues whose real part is negative and, respectively, positive. Analogous statements are known for perturbations of more complicated orbits. 

The simplest kind of an orbit is a fixed point, or an equilibrium. If a mechanical system is in a stable equilibrium state then a small push will result in a localized motion. In a system with damping, a stable equilibrium state is moreover asymptotically stable. On the other hand, for an unstable equilibrium, certain small pushes will result in a motion with a large amplitude that may or may not converge to the original state.
There are useful tests of stability for the case of a linear system. Stability of a non-linear system can often be inferred from the stability of its linearization.

In this paper, in section 2, the dynamics of the FLRW cosmology in
modified gravity is considered. Moreover, we discuss analytically the conditions for having
$\omega$ across over $-1$. In section
3, we study the numerical solution for a successful bouncing  and in section 4, we discuss about stability conditions of our model. In section 5, the cosmological parameters will be checked and finally, we summaries our paper in section 6.

\section{The Model}\label{Model}

As following our previous paper \cite{ghanaatian2014bouncing} we consider a $f(R,\phi)$ theory of gravity in the Einstein-Hilbert action which is replaced
by the square of the conformal Weyl tensor and matter lagrangian
\begin{eqnarray}\label{ac1}
I=-\frac{\alpha}{4}\int{d^4x\sqrt{-g}\left\{C_{\mu\nu\rho\lambda}C^{\mu\nu\rho\lambda}-g^{\mu\nu}\partial_{\mu}\phi\partial_{\nu}\phi
+2V(\phi)+2f(\phi)\mathcal{L}_m\right\}},
\end{eqnarray}
where $\alpha=1/8\pi G$, $G$ is the universal gravitational constant, the chameleon scalar field of $\phi$ is just dependent on cosmic time, $t$ and $V(\phi)$ is an arbitrary potential function dependent on $\phi$ and unlike the usual Einstein-Hilbert action, the matter Lagrangian density
$\mathcal{L}_m$ is modified by $f(\phi)\mathcal{L}_m$ and $C_{\mu\nu\rho\lambda}$ is the Weyl tensor as
\begin{eqnarray}\label{Weyl tensor}
C_{\mu\nu\rho\lambda}=R_{\mu\nu\lambda\rho}
-\frac{1}{2}(g_{\mu\lambda}R_{\nu\rho}-g_{\mu\rho}R_{\nu\lambda}-g_{\nu\lambda}R_{\mu\rho}+g_{\nu\rho}R_{\mu\lambda})
+\frac{R}{6}(g_{\mu\lambda}g_{\nu\rho}-g_{\mu\rho}g_{\nu\lambda})\cdot
\end{eqnarray}
The action (\ref{ac1}) when we put Planck units, i.e. $ c = \hbar = G = 1$, can be written as follows
\begin{eqnarray}\label{ac2}
I=-\frac{1}{32\pi}\int{d^{4}x\sqrt{-g}\left\{-g^{\mu\nu}\partial_{\mu}\phi\partial_{\nu}\phi
+R^{\mu\nu\rho\lambda}R_{\mu\nu\rho\lambda}-2R^{\mu\nu}R_{\mu\nu}+\frac{1}{3}R^{2}
+2V(\phi)+2f(\phi)\mathcal{L}_m\right\}}
\end{eqnarray}
since $\sqrt{-g}(R^{\mu\nu\rho\lambda}R_{\mu\nu\rho\lambda}-4R^{\mu\nu}R_{\mu\nu}+R^{2})$ is a total divergence (Gauss-Bonnet term), which doesn't contribute to the equation of motion and the action can be simplified as follows
\begin{eqnarray}\label{ac3}
I=-\frac{1}{16\pi}\int{d^{4}x\sqrt{-g}\left\{-\frac{1}{2}g^{\mu\nu}\partial_{\mu}\phi\partial_{\nu}\phi
+R_{\mu\nu}R^{\mu\nu}-\frac{1}{3}R^{2}+V(\phi)+f(\phi)\mathcal{L}_m\right\}}\cdot
\end{eqnarray}

Functional variation of the total action
with regard to the matter fields produces the equations of motion
while its functional variation considering the metric generates
the $f(R,\phi)$ modified gravity coupled by Weyl field equation.
Therefore, taking the variation of the action (\ref{ac3}) with
respect to the metric $g^{\mu\nu}$ , the field equations can be
obtained as {\cite{nojiri2006modified,starobinsky1980new,kerner1982cosmology,barrow1983stability,faraoni2006solar,schmidt2007fourth}}
\begin{eqnarray}\label{G}
R_{\mu\nu}-\frac{1}{2}R g_{\mu\nu}=8\pi\left( T_{\mu\nu}^{(R)}+T_{\mu\nu}^{(m)}\right)\cdot
\end{eqnarray}
where
\begin{eqnarray}\label{Tmunu}
8\pi T_{\mu\nu}^{(R)}=\partial_{\mu}\phi\partial_{\nu}\phi + g_{\mu\nu}(\frac{1}{2}g^{\alpha\beta}\partial_{\alpha}\phi\partial_{\beta}\phi
-V(\phi))+\mathcal{W}_{\mu\nu},
\end{eqnarray}
Also,
\begin{eqnarray}
\mathcal{W}_{\mu\nu}=&-&\frac{1}{2}g_{\mu\nu}\Box R-\Box R_{\mu\nu}+\nabla_{\rho}\nabla_{\mu}R_{\nu}^{\rho}+\nabla_{\rho}\nabla_{\nu}R_{\mu}^{\rho}-2R^{\rho}_{\mu}R_{\nu\rho}
+\frac{1}{2}g_{\mu\nu}R_{\rho\lambda}R^{\rho\lambda}\nonumber\\
&-&\frac{1}{3}(2\nabla_{\mu}\nabla_{\nu}R-2g_{\mu\nu}\Box R-2RR_{\mu\nu}+\frac{1}{2}g_{\mu\nu}R^2)\cdot
\end{eqnarray}

Here $R_{\mu\nu}$ is the Ricci tensor, $ T_{\mu\nu}^{(R)}=g_{\mu\nu}T_{\mu}^{\nu (R)}$  and $T_{\mu\nu}^{(m)}= \textrm{diag}(-\rho_m f, p_m f, p_m f, p_m f)$  is the modified energy-momentum tensor of the matter in the
prefect fluid form, respectively. The metric of the FLRW universe is given by
\begin{eqnarray}\label{FLRWmetric}
ds^2=-dt^2+a^2(t)\left(\frac{dr^2}{1-kr^2}+r^2d\theta^2+r^2 \sin^2\theta d\phi^2\right),
\end{eqnarray}
where $k=1,0,-1$ are for closed, flat and open geometries respectively. The first Friedman equation generality is
\begin{eqnarray}\label{Friedman1}
3H^2=8\pi \Sigma_{i}{\rho_i}-\frac{3k}{a^2},
\end{eqnarray}
that $H=\frac{\dot{a}}{a}$ is the Hubble parameter, the dot
denotes a derivative with respect to cosmic time $t$ and summation runs over the non-relativistic matter, radiation and
other components. The dark energy energy density, $\rho_X$, and the dark energy pressure, $p_X$, which can be obtained directly by metric is described by the holographic principle as,
\begin{eqnarray}\label{rho-pdark}
\rho_{X}&=&\frac{\alpha}{2-\alpha}\Omega_{m0}e^{-3N}+\beta e^{-\left(4-\frac{2}{\alpha}\right)N},\\
p_{X}&=&-\left(\frac{2}{3\alpha}-\frac{1}{3}\right) \beta e^{-\left(4-\frac{2}{\alpha}\right)N}\cdot
\end{eqnarray}
where $N=\ln{a}, $ $\Omega_{m0}$ is the relative density of the non-relativistic matter, the $\beta$ is an integration constant and $\alpha$ is a constant which are determined by \citep{gao2009holographic}. If the dark energy relative density be denoted by $\Omega_X$, the subscript $0$ denoted for all component's values of the present time (zero red-shift) and the equation state, $\omega_0\Omega_X$, of the dark energy be known, the value of $\alpha$ and $\beta$ can be determined by,
\begin{eqnarray}\label{alpha-beta}
\alpha&=&\frac{2\Omega_{X0}}{\Omega_{m0}+\Omega_{X0}-3\omega_0\Omega_{X0}},\nonumber\\
\beta&=&\frac{3\omega_0\Omega_{X0}^2}{3\omega_0\Omega_{X0}-\Omega_{m0}}
\cdot
\end{eqnarray}
So for current observations \citep{planck2015planck,ade2015planck}, and following values of parameters: $\Omega_{m0}=0.268$, $\Omega_{X0}=0.683$ and $\omega_0=-1$ we will have $\alpha=0.456$ and $\beta=0.603$ which will be used in our following calculations.
 
On the other hand, if we consider the spatially flat, $k=0$, FLRW metric for the universe, the set of field equations (\ref{G}) when $f\doteq f(\phi)$ reduce to the modified Friedmann equations in the framework of $f(R,\phi)$-modified gravity as
\begin{eqnarray}
3H^2 &=&8\pi\left( \rho_{R}+\rho_{m}f\right),\label{f1}\\
-2\dot{H}-3H^2&=&8\pi\left( p_{R}+p_{m}f\right),\label{f2}
\end{eqnarray}
where $\rho_R$ and $p_R$ are the Ricci curvature contributions of the energy density and pressure which have obtained from our action respectively. 
The model can be considered as a standard model with the effect of
the Weyl and $f(R,\phi)$ gravity modification contributed in the energy
density and pressure of the Friedman equations. Corresponding to standard spatially-flat
FLRW universe for the $00$ and $ii$ components yields,
 \begin{eqnarray}
8\pi \rho_{R}&=&\frac{1}{2}\dot{\phi}^2
+V(\phi)+\mathcal{W}_{00},\label{rho}\\
8\pi p_{R}&=&\frac{1}{2}\dot{\phi}^2-V(\phi)+\frac{\mathcal{W}_{ii}}{a^2(t)}\label{p}
\end{eqnarray}
where after some
algebraic calculation with $ R=6\dot{H}+12H^2$, we have
\begin{eqnarray}
\frac{\mathcal{W}_{00}}{3}\texttt{}&=&\dddot{H}(1-H)-4\ddot{H}H^2+2\dot{H}^2(1-2H-4H^2)-8H^4,\label{W00}\\
\frac{\mathcal{W}_{ii}}{a^2(t)}&=&4\ddot{H}(6H^2+H+3\dot{H})+\frac{3}{2}\dot{H}^2+\dot{H}\left(19H^2-12H
+\frac{9}{2}\right)+\frac{3}{2}\left(3+H^2\right)\cdot\label{Wii}
\end{eqnarray}

The variation of the action (\ref{ac3}) with respect to scalar field $\phi$ provides the wave equation for scalar field as,
\begin{eqnarray}
\ddot{\phi}+3H\dot{\phi}+V'+\gamma\rho_m f'=0\label{EOM},
\end{eqnarray}
where $\gamma=\frac{p_{m}}{\rho_{m}}$, the energy density $\rho_m$ stands for the contribution from the cold dark
matter to the energy density and prime indicated differentiation with respect to $\phi$.
The energy conservation laws are still given by
\begin{eqnarray}
\dot{\rho}_{m} +3H\rho_{m}(1 +\gamma)&=&-(1-\gamma)\rho_m\frac{\dot{f}}{f},\label{rhom}\\
\dot{\rho}_{R} +3H\rho_{R}(1 +\omega)&=&0,\label{rhoR}
\end{eqnarray}
and $\omega=\frac{p_{R}}{\rho_{R}}$ is the equation of state (EoS)
parameter due to the curvature contribution. Now, with integrating eq.(\ref{rhom}) respect to $t$ we have
\begin{eqnarray}
\rho_{m}=\frac{\rho_{0}}{f^{(1-\gamma)}a^{3(1+\gamma)}}\label{rhom1}
\end{eqnarray}
where $\rho_{0}$ is a constant of integration. By using eqs. (\ref{f1}), (\ref{f2}), (\ref{rho}), (\ref{p}) and in comparison with the standard Friedmann equations we identify $\rho_{eff}$ and $p_{eff}$ as
\begin{eqnarray}
\rho_{eff}&\doteq& \rho_{m} f+\frac{1}{8\pi}\left(\frac{1}{2}\dot{\phi}^2
+V(\phi)+\mathcal{W}_{00}\right)=\frac{3}{8\pi}H^2,\label{rhoeff}\\
p_{eff}&\doteq& \gamma\rho_{m}f+\frac{1}{8\pi}\left(\frac{1}{2}\dot{\phi}^2-V(\phi)+
\frac{\mathcal{W}_{ii}}{a^2(t)}\right)=-\frac{1}{8\pi}\left(2\dot{H}+3H^2\right)\label{peff}
\end{eqnarray}
 So for their effective equation of state parameter, $\omega_{eff}=\frac{p_{eff}}{\rho_{eff}}$, we will have
\begin{eqnarray}
\omega_{eff}&=&-1-\frac{2}{3}\frac{\dot{H}}{H^2}\nonumber\\
&=&-1+\frac{8\pi \rho_mf(1+\gamma)+\dddot{H}\mathcal{A}+\ddot{H}\mathcal{B}+\dot{H}^2\mathcal{C}+\dot{H}\mathcal{D}+\frac{3}{2}(3+H^2-16H^4)+\dot{\phi}^2}
{8 \pi\rho_mf+\frac{1}{2}\dot{\phi}^2+V(\phi)+\mathcal{W}_{00}},
\end{eqnarray}
where
\begin{eqnarray}
\mathcal{A}&\doteq&3(1-H),\,\,\,\,\,\,\,\,\,\,\,\,\,\,\,\,\,\,\,\,\,\,\,\,\,\,\,\,\,\,\,\,\,\,\
\mathcal{B}\doteq 4(3\dot{H}+3H^2+H),\nonumber\\
\mathcal{C}&\doteq&3(\frac{5}{2}-8H^2-4H),\,\,\,\,\,\,\,\,\,\,\,\
\mathcal{D} \doteq 19H^2-12H+\frac{9}{2}\cdot
\end{eqnarray}
In the absence  of scalar field, $\phi$ , from Eqs. (\ref{rhoeff}) and
(\ref{peff}), we have $\rho_{eff}= \mathcal{W}_{00},$ and $p_{eff}
= \frac{\mathcal{W}_{ii}}{a^2(t)}$. Therefore, these Eqs.
 don't transform to the usual Friedmann
equations in GR
 and it's seriously dependent on functionality of scale factor respect to cosmic time, $t$.
However, from these equations and Eqs. (\ref{rhom}) and (\ref{rhoR}), we could
obtain $\dot{H} > 0$ and $\dot{H} < 0$ for the phantom,
 $\omega_{eff} <-1$, and quintessence, $\omega_{eff} >-1$, respectively. So, we need to probe more into scale factor, $a(t)$, as the following sections.

At this stage, the cosmological evolution of EoS parameter,
$\omega_{eff}$, is considered, and we show analytically there are some conditions that cause the EoS parameter
cross the phantom divide line ($\omega_{eff}\rightarrow -1$). To do that, $\rho_{eff} + p_{eff}$ must be disappeared
at the bouncing point. For investigating
this possibility, we have to check the condition
$\frac{d}{dt}(\rho_{eff}+p_{eff})\neq 0$ when $\omega_{eff}\rightarrow -1$. Using Eqs. (\ref{W00}), (\ref{Wii}) in (\ref{rhoeff}) and (\ref{peff}) we have,
\begin{eqnarray}\label{rho plus p}
\frac{d}{dt}(\rho_{eff}+p_{eff})=-3(H-1)\ddddot{H}+\mathcal{E}\dddot{H}+12\ddot{H}^{2}+\mathcal{F}\ddot{H}+\mathcal{G}\dot{H}^3+\mathcal{J}\dot{H}^2+\mathcal{K}\dot{H}+\mathcal{O}\neq 0,
\end{eqnarray}
where
\begin{eqnarray}
\mathcal{E}&\doteq&12H^{2}+9\dot{H}+4H,\,\,\,\,\,\,\,\,\,\,\,\,\,\,\
\mathcal{F}\doteq -(48\dot{H}H^2-19H^2+12H-19\dot{H}-\frac{9}{2})\nonumber\\
\mathcal{G}&\doteq& -12(4H+1),\,\,\,\,\,\,\,\,\,\,\,\,\,\,\,\,\,\,\,\,\,\,\,\,\,\, \mathcal{J}\doteq 2(19H-6),\,\,\,\,\,\,\,\,\,\,\,\,\,\,\,\ \mathcal{K}\doteq -3(32H^3-H)\nonumber\\
\mathcal{O}&\doteq& 2\dot{\phi}\ddot{\phi}+8\pi\rho_m(\gamma+1)\dot{f}\cdot
\end{eqnarray}

In this case, our analytical discussion about $\omega_{eff}\rightarrow
-1$ would be just a boring game on different components of
Eq.(\ref{rho plus p}) to satisfy $\frac{d}{dt}(\rho_{eff}+p_{eff})\neq 0$.
For example, if second and upper orders of derivatives of $H$
respect to cosmic time $t$ vanish, one can find,
\begin{eqnarray}
\dot{H} &\neq& \frac{1}{3\mathcal{G}}\left(\frac{\mathcal{S}}{2}-2(\frac{3\mathcal{G}\mathcal{K}-\mathcal{J}^2}{\mathcal{S}})+\mathcal{J}\right),
\end{eqnarray}
or
\begin{eqnarray}
\dot{H} &\neq& -\frac{1}{3\mathcal{G}}\left(\frac{\mathcal{S}}{4}-(\frac{3\mathcal{G}\mathcal{K}-\mathcal{J}^2}{\mathcal{S}})-\mathcal{J}\right)\pm \frac{\sqrt{3}}{6\mathcal{G}}i\left(\frac{\mathcal{S}}{2}+2(\frac{3\mathcal{G}\mathcal{K}-\mathcal{J}^2}{\mathcal{S}})\right),
\end{eqnarray}
where
\begin{eqnarray}
\mathcal{S}&\doteq&\sqrt [3]{108\mathcal{O}\mathcal{G}^2-36\mathcal{GJK}+12\sqrt{3}\mathcal{S'}\mathcal{G}+8\mathcal{J}^3}\nonumber\\
\mathcal{S}'&\doteq&\sqrt{27\mathcal{G}^2\mathcal{O}^2-18\mathcal{GJKO}+4\mathcal{GK}^3+4\mathcal{K}^3\mathcal{O}-\mathcal{J}^2\mathcal{K}^2}
\end{eqnarray}

\section{Numerical solution for a successful bouncing}\label{sec:NSSB}
In FLRW cosmological model, for only time dependent $\phi$ by
invariance of the action under changing fields and vanishing
variations at the boundary, the equations of motion for scalar
fields $\phi$ can be rewritten by
\begin{eqnarray}
\ddot{\phi}+3H\dot{\phi}+\delta_2 V_0e^{\delta_2\phi}+\gamma\delta_1\rho_mf_0e^{\delta_1\phi}=0
\end{eqnarray}
when we define  $f\doteq f(\phi)=f_0e^{\delta_1\phi}$ and $V\doteq V(\phi)=V_0e^{\delta_2\phi}$ with $\delta_1$ and $\delta_2$ which are dimensionless constants.

Also the solution for $H(t)$, Eq. (\ref{f1}), provides a dynamical
 universe with contraction for $t<0$, bouncing at $t=0$ and then expansion
  for $t>0$. The above analysis can be clearly seen in the numerical calculation given in Fig. 1.\\
\begin{tabular*}{2.5 cm}{cc}
\includegraphics[scale=.35]{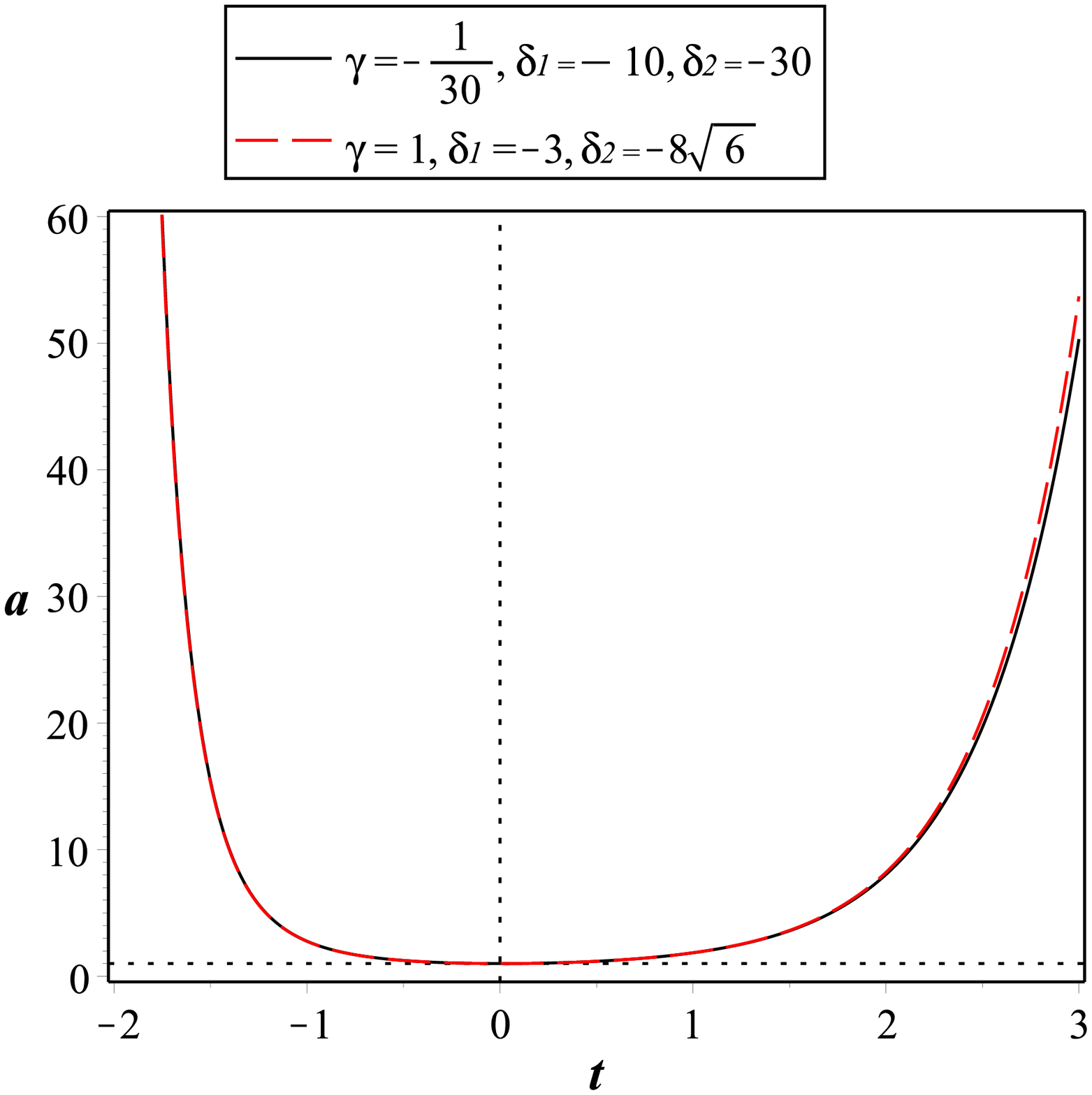}\hspace{1 cm}\includegraphics[scale=.35]{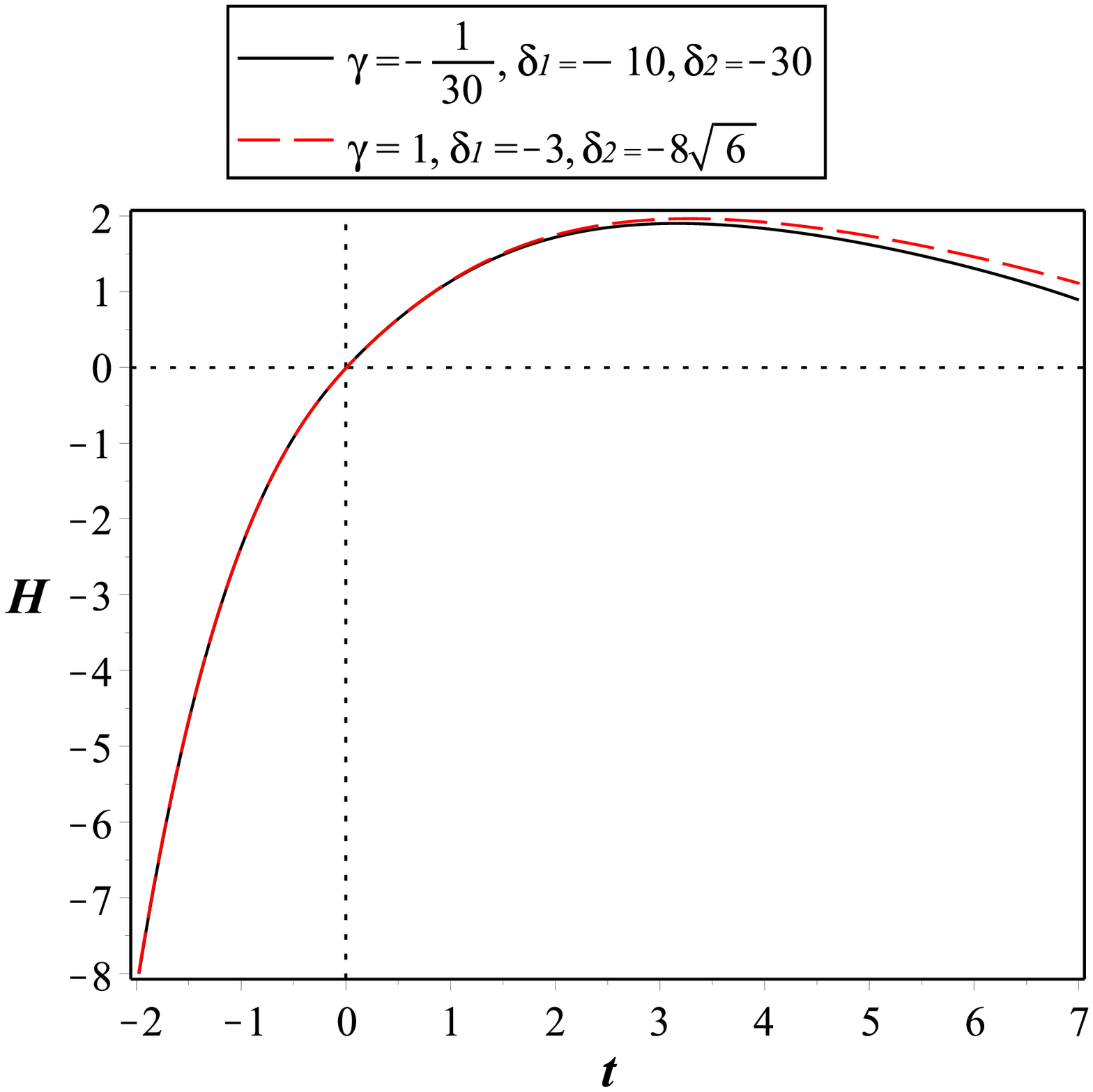}\hspace{1 cm}\\
\hspace{1 cm} Fig.1: \, The graph of scalar factor, $a$, and $H$,  plotted as function of time,\\
\hspace{1 cm} for $f_0 = -2$, $V_0 = 1$, $\rho_m= 0.3$, with initial values as,\\
\hspace{1 cm} $\phi(0) = 0.5$, $\dot{\phi}(0) = -0.1$, $\dot{a}(0)=-0.01$ and $\ddot{a}(0) = 1.5$\\
\end{tabular*}\\
\\
\begin{tabular*}{2.5 cm}{cc}
\includegraphics[scale=.35]{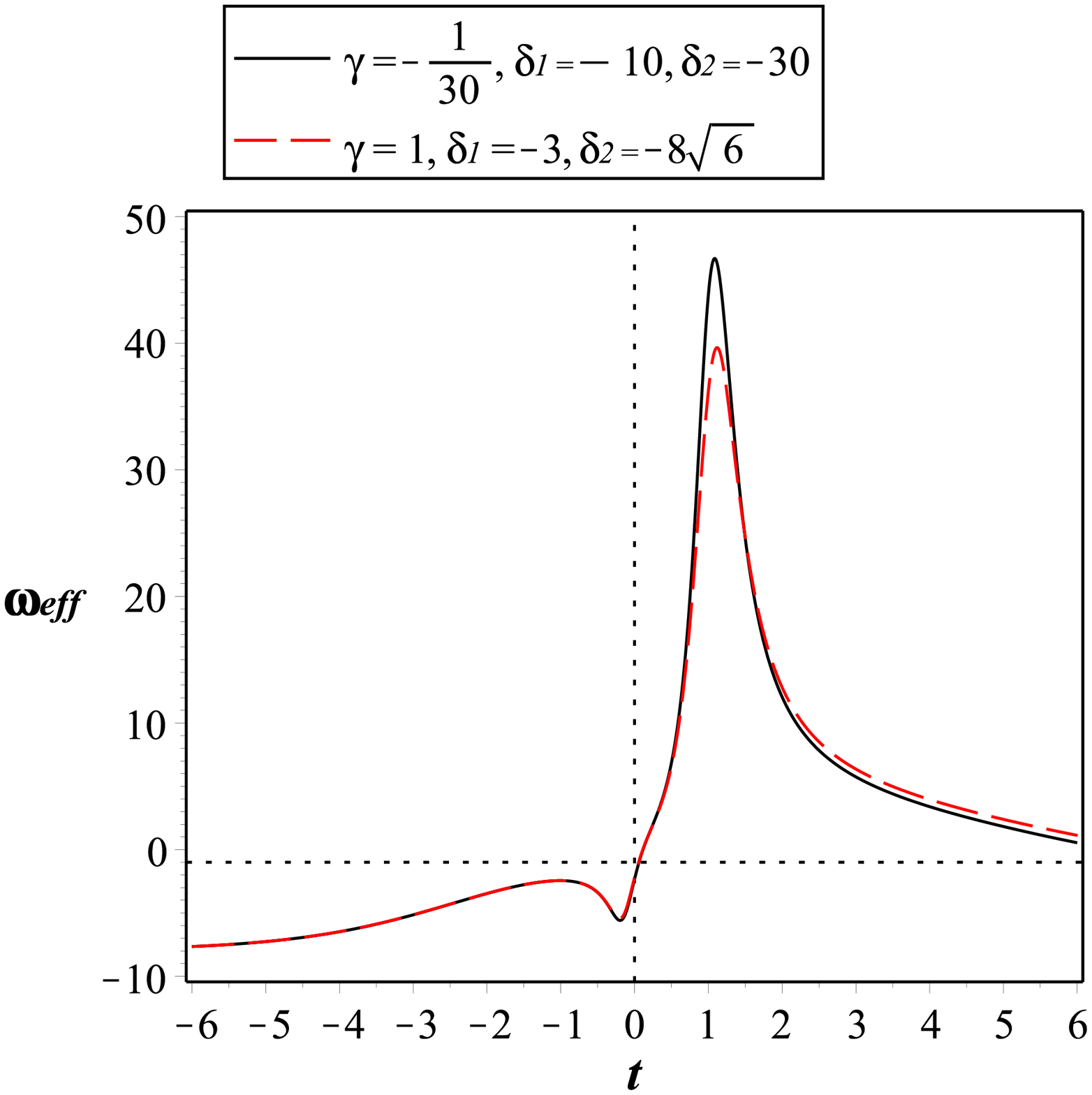}\hspace{1 cm}\includegraphics[scale=.35]{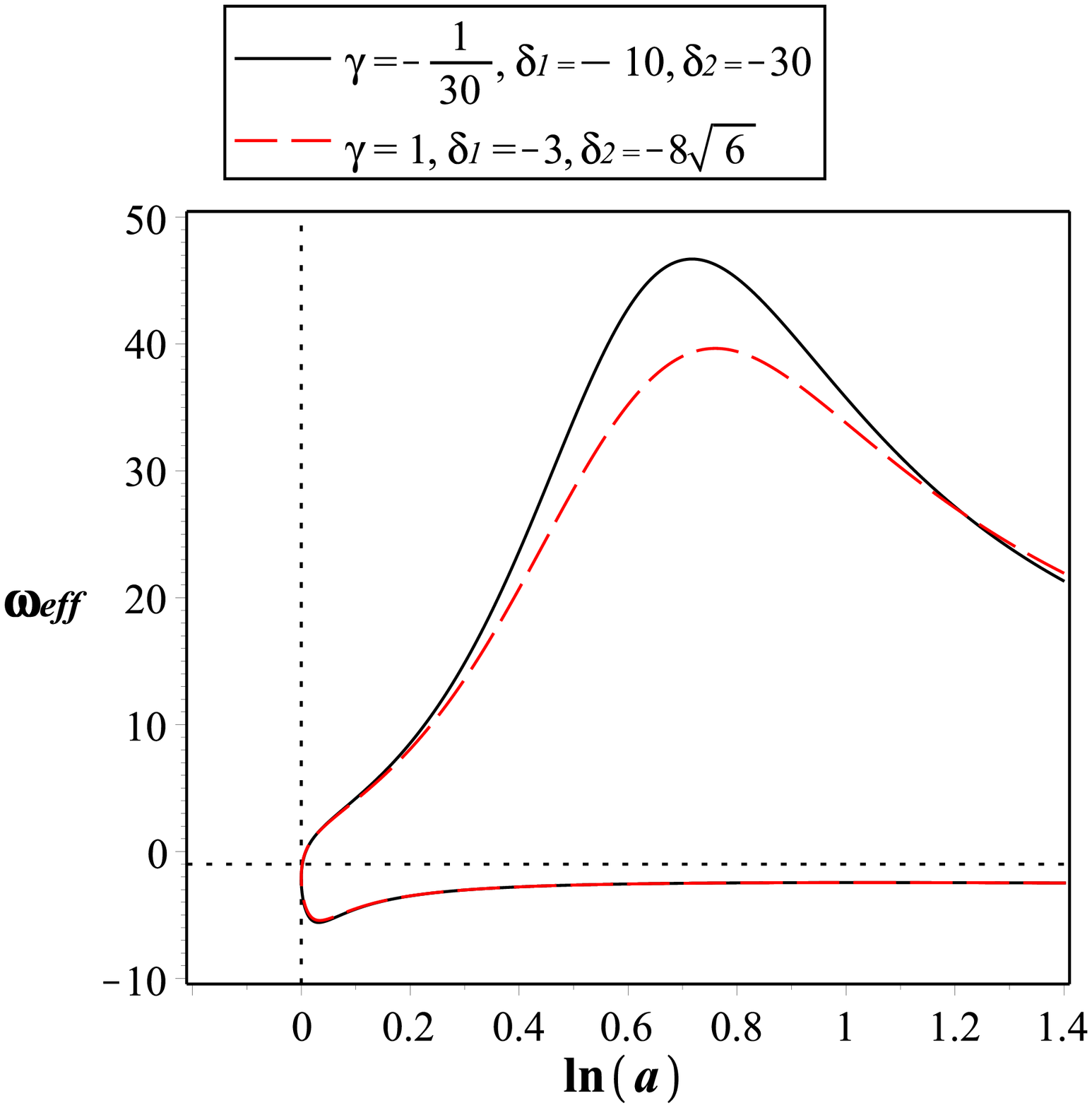}\hspace{4 cm}\\
\hspace{1 cm} Fig.2: \, Plot of the evolution of the $\omega_{eff}$ as a function of cosmic time $t$ and $\ln(a)$,\\
\hspace{1 cm} for $f_0 = -2$, $V_0 = 1$, $\rho_m= 0.3$, with initial values as,\\
\hspace{1 cm} $\phi(0) = 0.5$, $\dot{\phi}(0) = -0.1$, $\dot{a}(0)=-0.01$ and $\ddot{a}(0) = 1.5$\\
\end{tabular*}\\

In our model, the EoS parameter crosses $-1$
line from $\omega_{eff}<-1$ to $\omega_{eff}>-1$, as Fig.2, which is supported
by observations. This model bears the same as quintom
dark energy models which includes two quintessence and phantom
fields. For a successful bounce implying, a list of
test on the necessary conditions is needed that during the
contracting phase, the scale factor $a(t)$ should be decreased,
i.e., $\dot{a} < 0$, and in the expanding phase, we should have
$\dot{a} > 0$. At the bouncing point, $\dot{a} = 0$, and so around
this point $\ddot{a} > 0 $ for a period of time, the Hubble
parameter $H$ runs across zero from $H < 0$ to $H > 0$ and $H = 0$
at the bouncing point. According to Fig.1 and Fig.2, at $t\rightarrow 0$, $\omega_{eff}<-1$ and
$\dot{H}$ are positive and we see that at the bouncing point where
the scale factor $a(t)$ is not zero, we avoid singularity faced in
the usual Einstein cosmology.

\section{Stability conditions}\label{sec:StabCon}

In this section, we consider the structure of the dynamical system via phase plane analysis,
by introducing the following dimensionless variables,
\begin{eqnarray}\label{dimensionless variables}
x=\frac{\dot{\phi}}{\sqrt{6}H}\,\,\,\,\,\,\,\,y=\frac{\rho_mf(\phi)}{3H^2}\,\,\,\,\,\,\,\, z=\frac{V(\phi)}{3H^2},\,\,\,\,\,\,\,\,w_{0}=\frac{\mathcal{W}_{00}}{3H^2},\,\,\,\,\,\,\,\,  \textrm{and} \,\,\,\,\,\,\,\,w_{i}=\frac{\mathcal{W}_{ii}}{3\dot{a}^2},\,\,\,\,\,\,\,\,
\end{eqnarray}
With combining eqs.(\ref{f1}), (\ref{rho}) and (\ref{dimensionless variables}) we have
\begin{eqnarray}\label{st1}
w_{0}=1-(x^2+y+z)
\end{eqnarray}
where
\begin{eqnarray}\label{St H1pH2}
\frac{\dot{H}}{H^2}=-\frac{3}{2}[x^2+\gamma y-z+w_i+1]\label{H1H2}
\end{eqnarray}
Now with previous section definitions, $f\doteq f(\phi)=f_0e^{\delta_1\phi}$ and $V\doteq V(\phi)=V_0e^{\delta_2\phi}$, characterizing the slope of potential $V(\phi)$ and $f(\phi)$, eq.(\ref{EOM}) can be rewritten as
\begin{eqnarray}
\ddot{\phi}=-3H\dot{\phi}-\delta_2 V-\gamma\delta_1\rho_mf
\end{eqnarray}
and we can find the evolution equations of these variables as
\begin{eqnarray}
x'&\doteq&\frac{dx}{dN}=-\left[3x+\frac{\sqrt{6}}{2}\delta_1\gamma y+\frac{\sqrt{6}}{2}\delta_2z
+x\frac{\dot{H}}{H^2}\right],\label{x'1}\\
y'&\doteq&\frac{dy}{dN}=\sqrt{6}\delta_1 xy-2y\frac{\dot{H}}{H^2},\label{y'1}\\
z'&\doteq&\frac{dz}{dN}=\sqrt{6}\delta_2 xz-2z\frac{\dot{H}}{H^2},\label{z'1}\\
w'_0&\doteq&\frac{dw_0}{dN}=6x^2-
\sqrt{6}\delta_1xy(1-\gamma)+2(1-w_0)\frac{\dot{H}}{H^2},\label{w0'1}\\
w'_i&\doteq&\frac{dw_i}{dN}= 6x^2+2\sqrt{6}\delta_2xz-2(1+w_i)\frac{\dot{H}}{H^2}-\frac{2}{3}\frac{\ddot{H}}{H^3}\label{wi'1}
\end{eqnarray}
where $N=\ln a$. It's clear the forth term of eq.(\ref{wi'1}) is not dependent on scalar field $\phi$ and it can be a big trouble for us, so we use a trick to omit it. We know
\begin{eqnarray}
\frac{\ddot{H}}{H^3}=\frac{d(\frac{\dot{H}}{H^2})}{dN}+2(\frac{\dot{H}}{H^2})^2\label{H2H3},
\end{eqnarray}
then, by using eq.(\ref{H1H2}) and eqs. (\ref{x'1}-\ref{wi'1}) one can find this term will be removed from both sides of our equation simply and by solving the new equation we obtain
\begin{eqnarray}
{\it w_i}={\frac { \sqrt {6}
}{3}}\,k- \left( {x}^{2}+\gamma\,y-z-1 \right)
\label{wi},
\end{eqnarray}
where
\begin{eqnarray}
k\doteq {\frac {\left( \delta_1\,\gamma\,y+\delta_2\,z \right)}{x}},\label{k}
\end{eqnarray}
 and in the following, eq.(\ref{wi'1}) can be rewrited as
\begin{eqnarray}
w'_i&=& 6\,x^2+3\,k^2
+\sqrt{6}\,k+\,\frac{\sqrt{6}}{x}\,\delta_2\,z
+2(\delta_1^2\gamma y +\delta_2^2z)
+2\sqrt{6}\delta_2xz\nonumber\\
&-&\left[\frac{\sqrt{6}}{3}\,k
\left(2+\frac{1}{x}\right)-2(x^2+\gamma y -z)
\right]
\frac{\dot{H}}{H^2}\label{wi'3}
\end{eqnarray}
replacing eq. (\ref{St H1pH2}) in above equations gives
\begin{eqnarray}
x'&=&-3x,\label{x'3}\\
y'&=&\sqrt{6}\,\left(\delta_1 x+k\right)y,\label{y'3}\\
z'&=&\sqrt{6}\,\left(\delta_2 x+k\right)z,\label{z'3}\\
w'_0&=&6x^2-(\delta_1 y+\delta_2  z)x-\sqrt{6}\,k\,(y+z),\label{w0'3}\\
w'_i&=&6x^{2}+2\left( {\delta_1}^{2}\gamma y + {\delta_2}^{2}z
   \right)-\sqrt {6}\,k\,\left(x^{2}+ \gamma\,y
  -z -1\right)
+
k ^{2}\,\left(5+ \frac{1}{x}\right)
\label{wi'5}
\end{eqnarray}
which by using the constraints (\ref{st1}) and (\ref{wi}), the eqs. (\ref{x'3})-(\ref{wi'5}) finally will been reduced to
\begin{eqnarray}
x'&=&\left( \delta_1\,y+\delta_2\,z \right)\left(\frac{1-\sqrt {6}}{2}\right)-3\,x\label{x'4}\\
y'&=&\sqrt{6}\,\left(\delta_1 x+k\right)y\label{y'4}\\
z'&=&\sqrt{6}\,\left(\delta_2 x+k\right)y\label{z'4}
\end{eqnarray}
Now with solving the system of eqs. (\ref{y'1})-(\ref{z'1}) and for scape than singularity at $t=0$ we must have $a_0\doteq a(0)\neq 0$ for example $a_0=1$ that it shows us the necessity of using the other theories except big bang theory as big bouncing theory. In this way we can find
\begin{eqnarray}
\ln (\frac{a}{a_0}) &=& \frac{1}{\sqrt{6}(\delta_2-\delta_1)x}\ln (\frac{z}{y}), \label{lna}\\
\textrm{for}\,\,\,\,\ a_0=1\,\,\,\,\,\rightarrow\,\,\,\,H &=& \frac{1}{\sqrt{6}(\delta_2-\delta_1)x}\left[\frac{y\dot{z}
-z\dot{y}}{yz}-\frac{\dot{x}\ln(\frac{z}{y})}{x}\right]\cdot\label{H}
\end{eqnarray}

Now for investigating the properties of the dynamical system we can obtain the critical points and study the stability of these
points. Critical points are always exact constant solutions in the context of autonomous
dynamical systems. These points are often the extreme points of the orbits and therefore
describe the asymptotic behavior. In the following we can find two fixed points ( or critical
points) by simultaneously solving $\frac{dx}{dN} = 0$, $\frac{dy}{dN} = 0$,  and $\frac{dz}{dN} = 0$. Substituting linear perturbations $x'\rightarrow x'+\delta x'$, $y'\rightarrow y'+\delta y'$ and $z'\rightarrow z'+\delta z'$ about the critical points into the three independent equations, to the
first orders in the perturbations, which yield three eigenvalues $\lambda_i$, $\lambda_j$ and $\lambda_k$ . Stability requires the
real part of all eigenvalues to be negative. The two fixed points depend on the different
values of $\gamma$, $\delta_1$ and $\delta_2$, as illustrated in Table 1.

\begin{table}[ht]
\caption{Critical points} % title of Table
\centering % used for centering table
\begin{tabular}{c c c c} % centered columns (4 columns)
\hline\hline %inserts double horizontal lines
$\textbf{Points}$ & $\textbf{x}$ & $\textbf{y}$ & $\textbf{z}$ \\ [0.5ex] % inserts table
%heading
\hline % inserts single horizontal line
$\textbf{P}_\textbf{1}\textbf{:}$ & $\frac{1}{5}\frac{6+6\sqrt{6}}{\delta_2}$ & 0 & $-\frac{36}{25}\frac{7+2\sqrt{6}}{\delta^2_2}$\\ [1ex]% inserting body of the table
$\textbf{P}_\textbf{2}\textbf{:}$ & $-\frac{6\gamma(\sqrt{6}-1)}{\delta_1(2\sqrt{6}-7)}$ & $\frac{36\gamma}{\delta_1^2(2\sqrt{6}-7)}$ & 0 \\ [1ex] % [1ex] adds vertical space
\hline %inserts single line
\end{tabular}
\label{table:nonlin} % is used to refer this table in the text
\end{table}

\subsection{Stability conditions for $\gamma=0$}\label{subsec:SCGamma0}
In case of $\gamma = 0$ we find two fixed points which for first point, we obtain three eigenvalues $\lambda_i$ and $i = 1, 2, 3$.  But for second point, the coordinate of critical point is (0 , 0 , 0) and we can't obtain any eigenvalue so this point will be unstable. In this way for the first point, $P_1$ we have
\begin{eqnarray}
(x_c\,,\, y_c\,,\, z_c)&=&(x_c\,,\, 0\,,\, -x_c^2)= \left( \frac{6}{5\delta_2}(1+\sqrt{6})\,,\,\, 0 \,\, ,\,-\left(\frac{6}{5\delta_2}(1+\sqrt{6})\right)^2\right),\,\,\,\,\nonumber\\
\lambda_{1\,P_1}&=& -{\frac {36\,\left(\delta_1-\delta_2\right)}{\delta_2\,
 \left( -6+\sqrt {6} \right) }}\nonumber\\
\lambda_{2\,P_1}&=& -\frac{3}{5}\left(\sqrt {6}+{
\frac {17}{2}}-\frac{1}{2}\,\sqrt {553+108\,\sqrt {6}}\right)\simeq +2.01\nonumber\\
\lambda_{2\,P_1}&=& -\frac{3}{5}\left(\sqrt {6}+{
\frac {17}{2}}+\frac{1}{2}\,\sqrt {553+108\,\sqrt {6}}\right)\simeq -15.15
\end{eqnarray}
It's clear the first point, $P_1$, is unstable too, because $\lambda_{2\,P_1}$ has a positive value. Therefore because all coordinates of $P_1$ is independent on $\gamma$ then this point will remain unstable in the all conditions and our investigations will been limited to second point.

\subsection{Stability conditions for second fixed point}\label{subsec:SCSFP}
The eigenvalues of second critical point generally obtained by
\begin{eqnarray}
\lambda_{1P_2}&=&36\,\gamma\,\frac{\delta_1-\delta_2}{\delta_1\,(\sqrt{6}-6)}\nonumber\\
^{\lambda_{2P_2}}_{\lambda_{3P_2}}&=&-\frac{3}{5}\,\gamma\,(\sqrt{6}+6)-\frac{3}{2}\pm\frac{3}{10}\sqrt{24\,\gamma^2(2\,\sqrt{6}+7)+60\,\gamma\,(\sqrt{6}+6)+25}
\end{eqnarray}
Therefore the suitable conditions for a successful stability will be happened when we have $ -\frac{1}{3}<-0.0507642711\preceq \gamma<0$ and $\delta_2> \delta_1$ if $\delta_1>0$ or $\delta_2< \delta_1$ if $\delta_1<0$. In this way asymptotically $z= -1$ stable equilibrium sink for 8 different classified parameters  while $y$ is varied respect of $x$, has been shown by figures 3 to 7.
\\
\begin{tabular*}{2.5 cm}{cc}
\includegraphics[scale=.4]{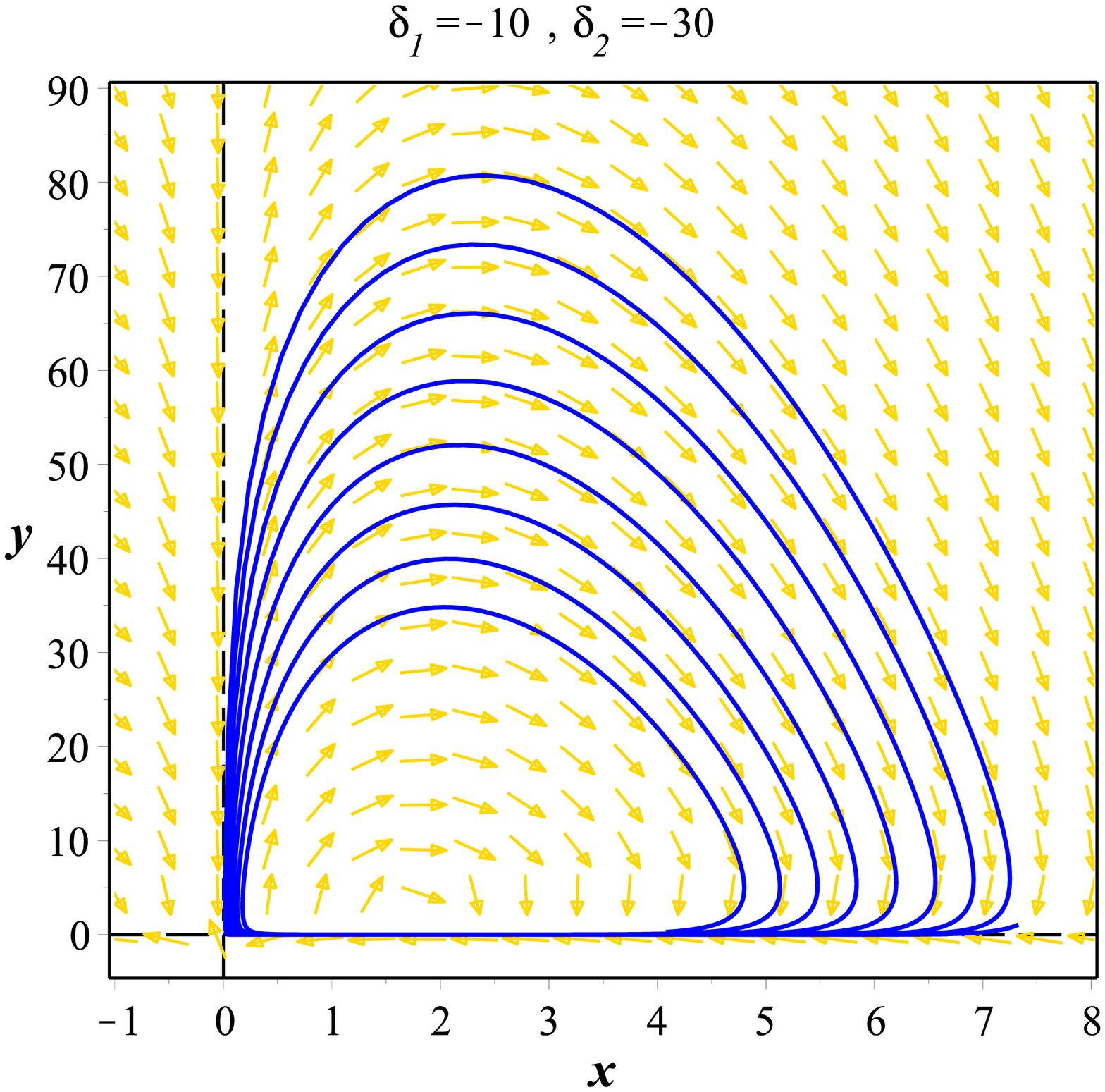}\hspace{0.5 cm}\includegraphics[scale=.4]{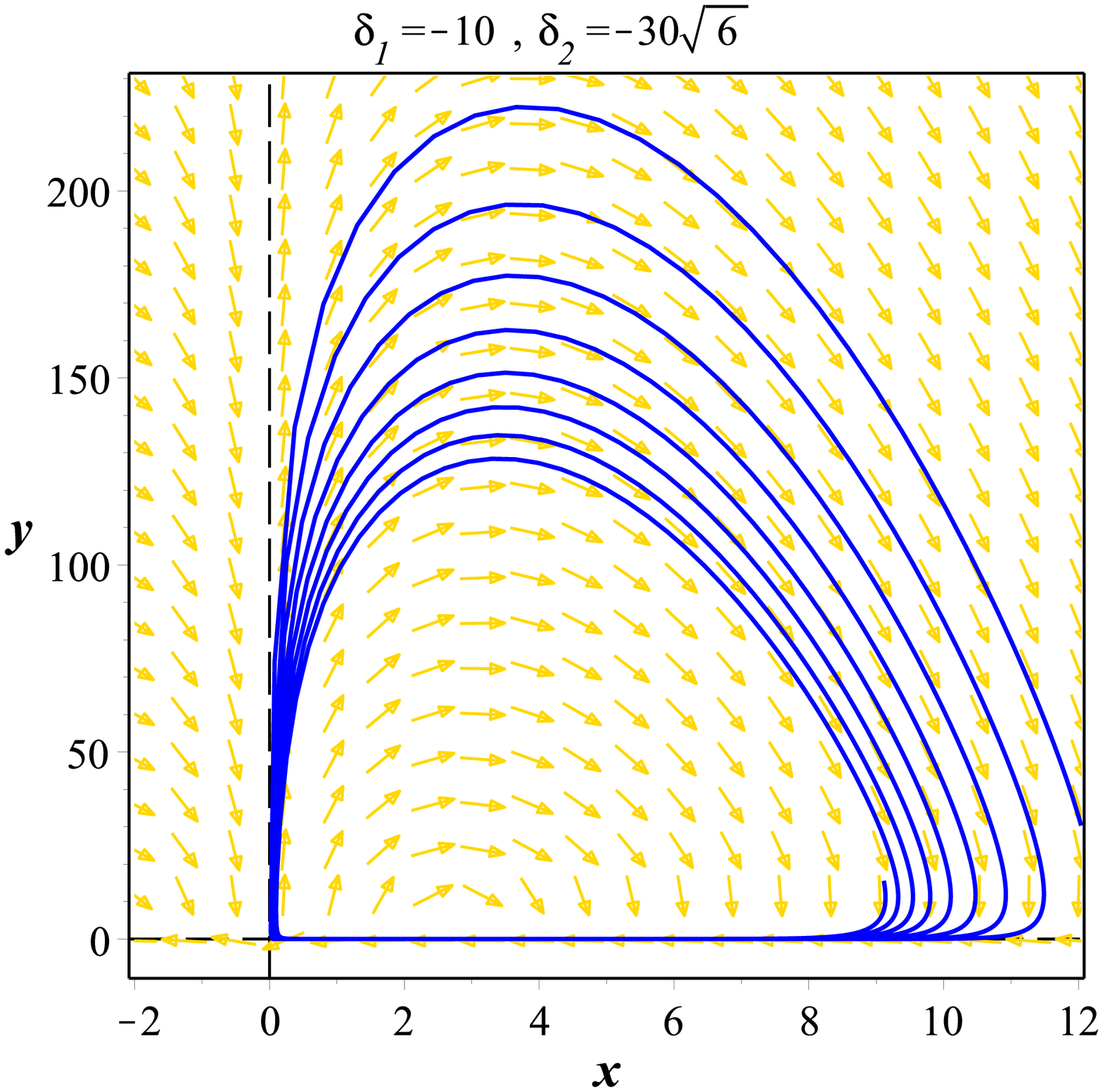}\\
\hspace{1 cm} Fig.3: \, Asymptotically $z= -1$ stable equilibrium sink for $\gamma=-\frac{1}{30}$\\
\hspace{1 cm}  with initial values which have been considered by $ x(0) = 0.1$ and $y(0) = 0.07$.\\
\end{tabular*}\\
\\
\begin{tabular*}{2.5 cm}{cc}
\includegraphics[scale=.4]{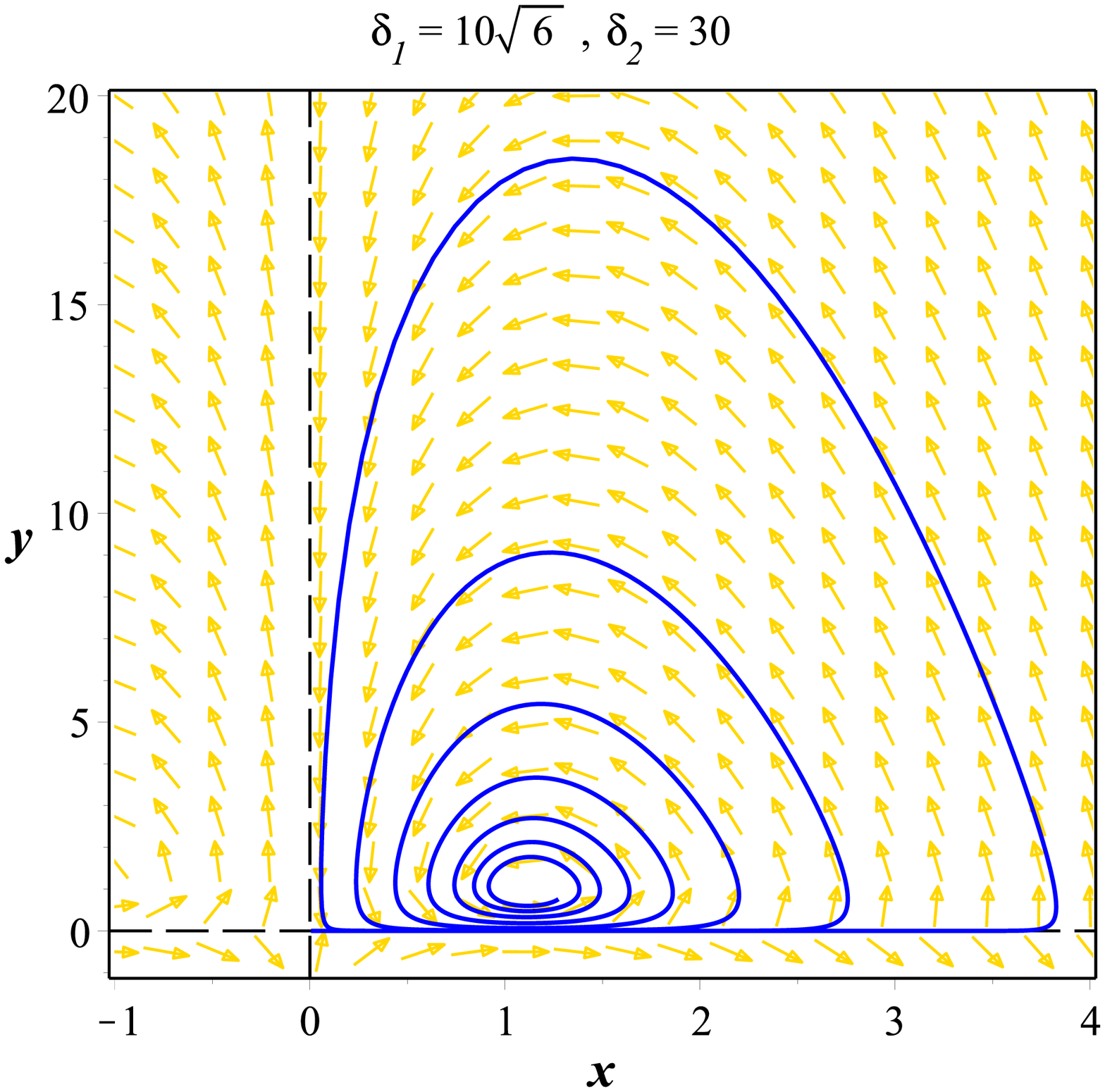}\hspace{0.5 cm}\includegraphics[scale=.4]{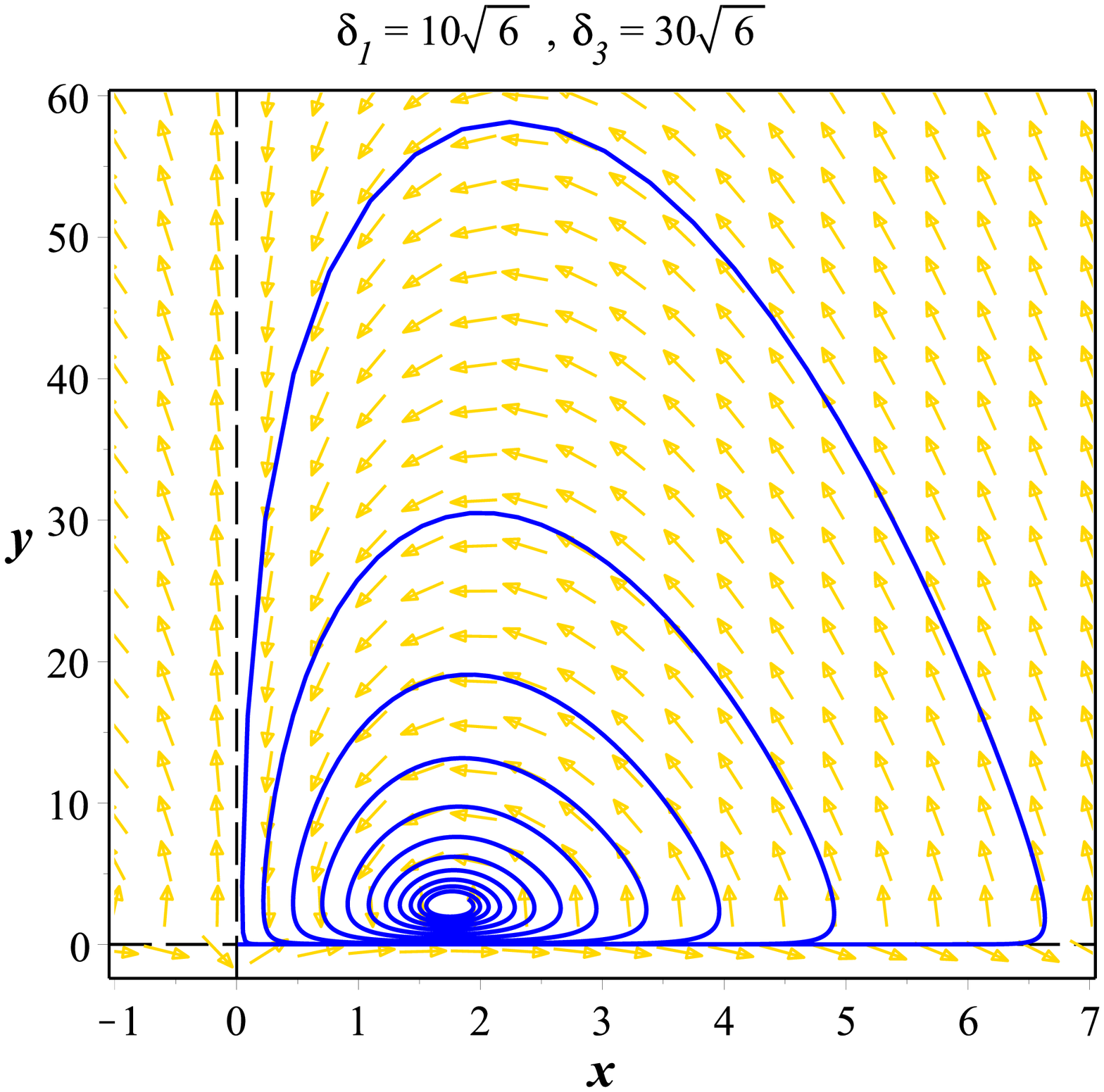}\\
\hspace{1 cm} Fig.4: \, Asymptotically $z= -1$ stable equilibrium sink for $\gamma=-\frac{1}{30}$\\
\hspace{1 cm}  with initial values which have been considered by $ x(0) = 0.1$ and $y(0) = 0.07$.\\
\end{tabular*}\\
\\
\begin{tabular*}{2.5 cm}{cc}
\includegraphics[scale=.4]{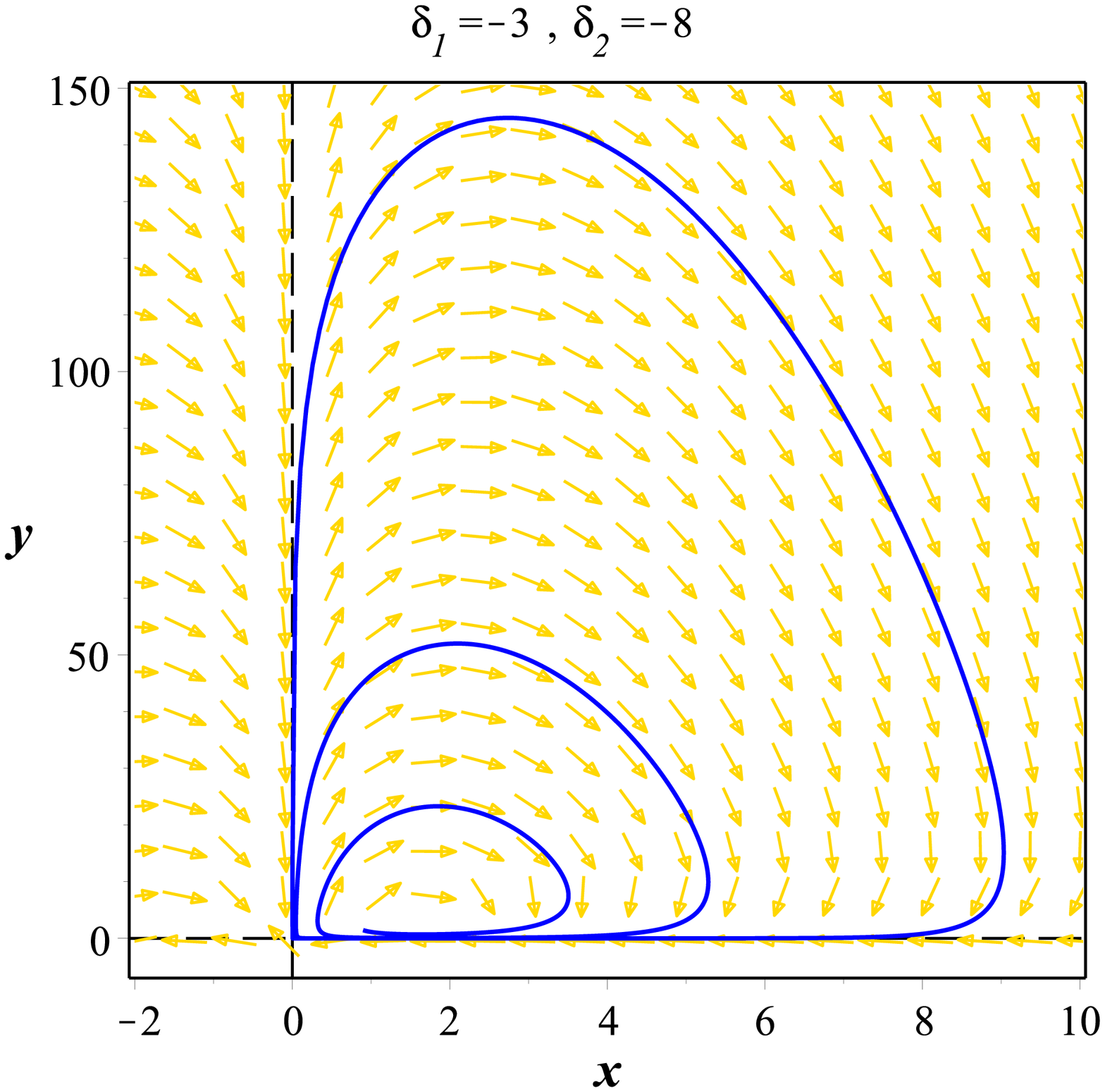}\hspace{0.5 cm}\includegraphics[scale=.4]{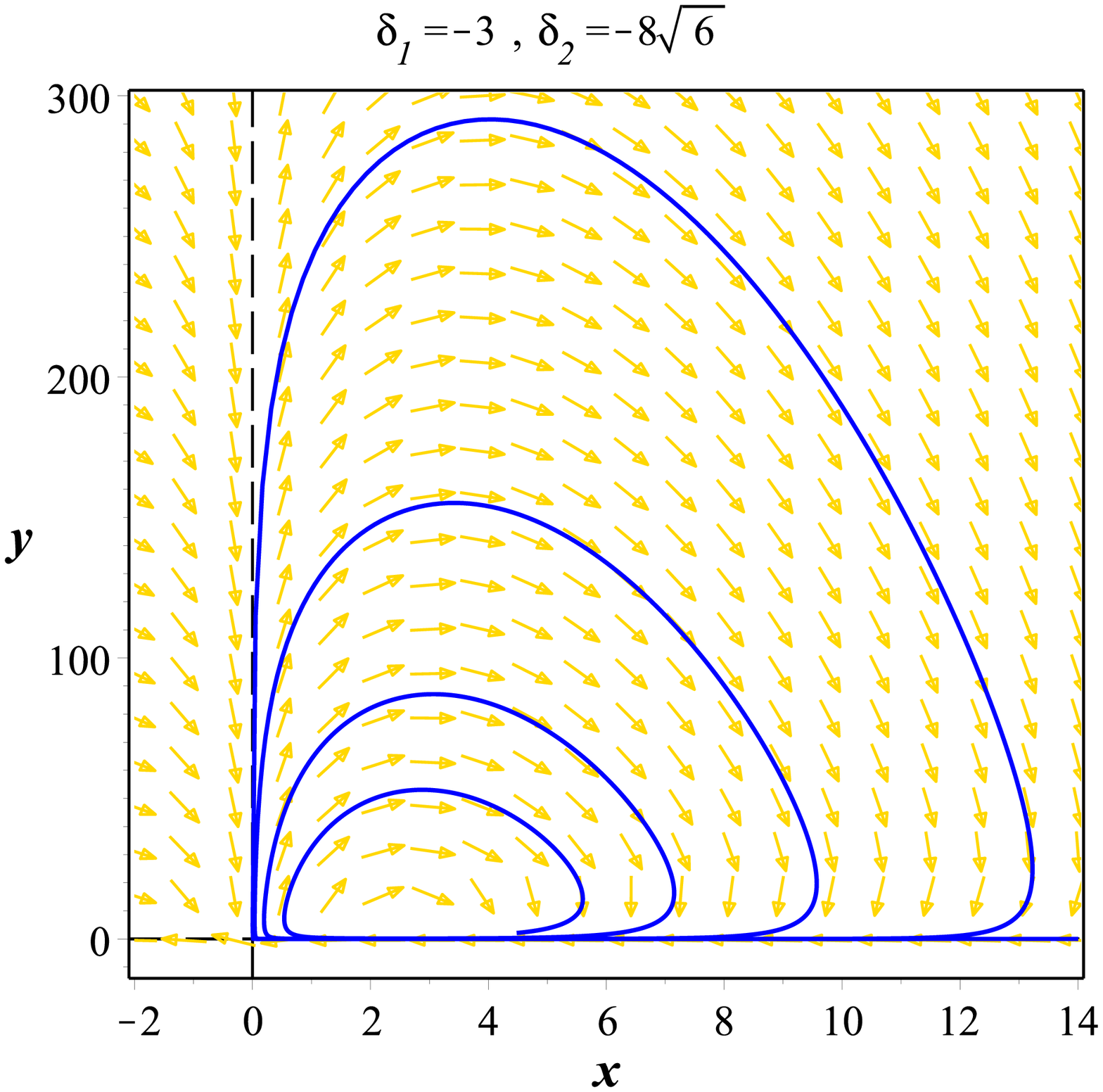}\\
\hspace{1 cm} Fig.5: \, Asymptotically $z= -1$ stable equilibrium sink for $\gamma=-\frac{1}{30}$\\
\hspace{1 cm}  with initial values which have been considered by $ x(0) = 0.1$ and $y(0) = 0.07$.\\
\end{tabular*}\\
\\
\begin{tabular*}{2.5 cm}{cc}
\includegraphics[scale=.4]{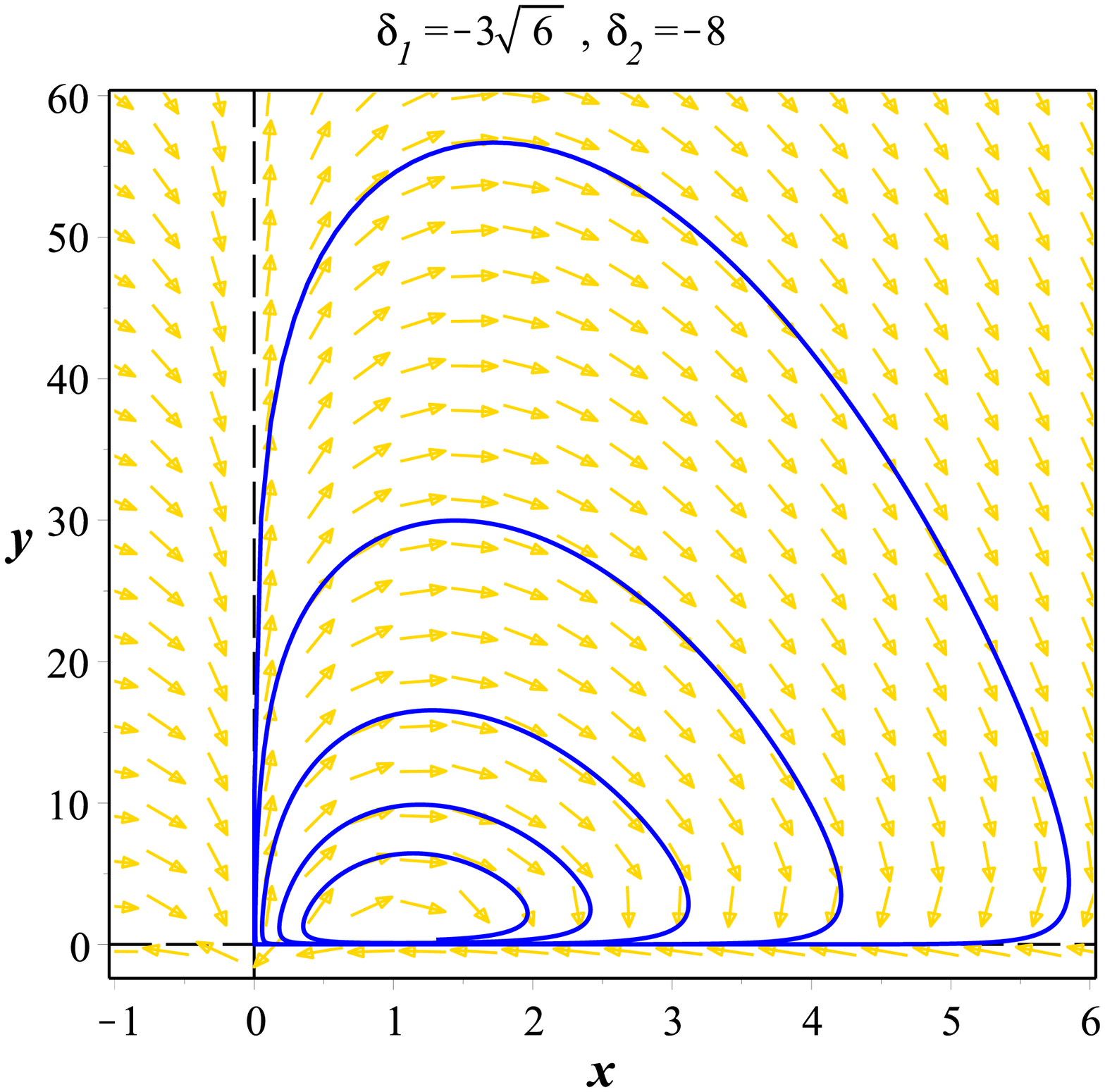}\hspace{0.5 cm}\includegraphics[scale=.4]{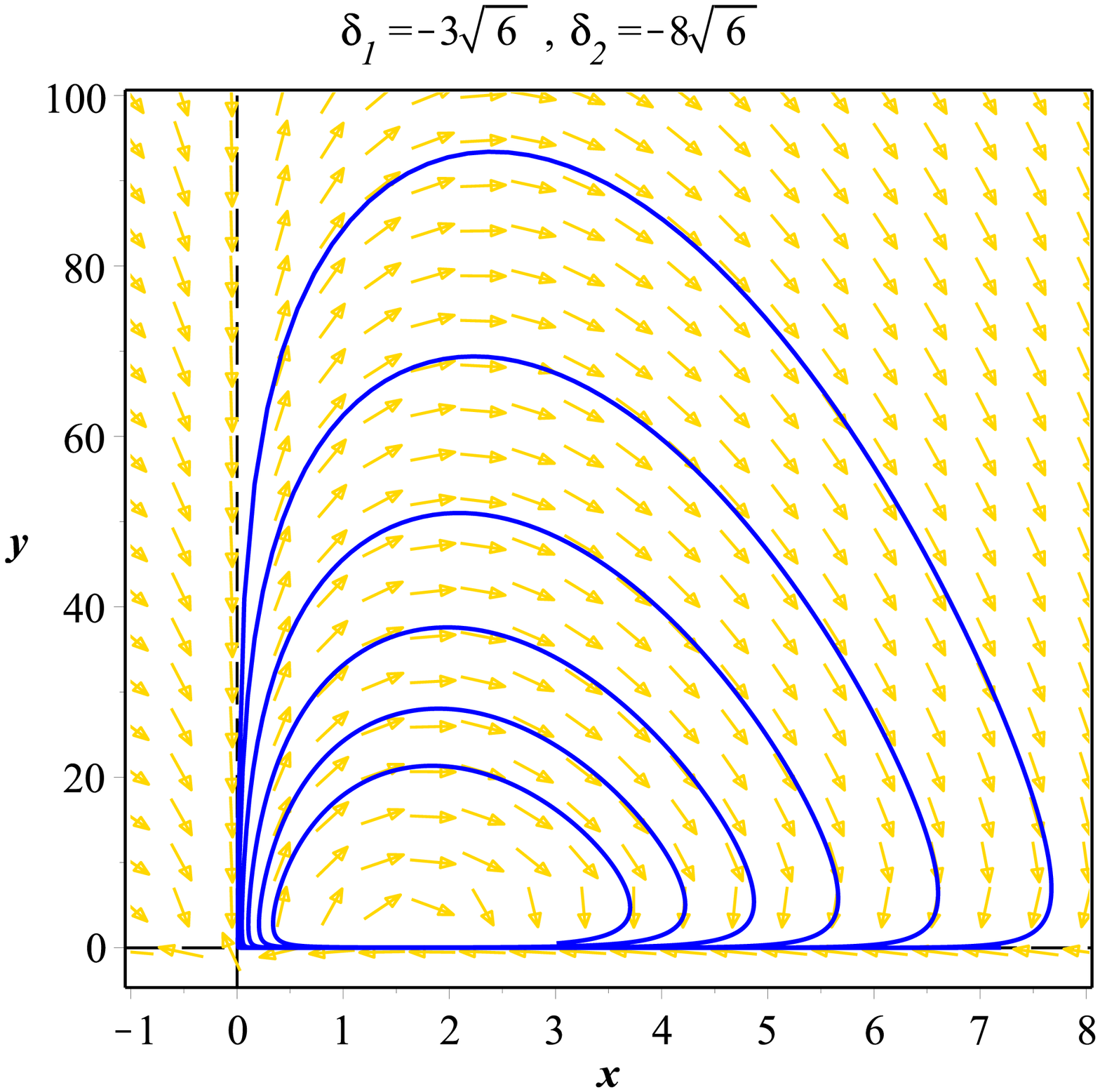}\\
\hspace{1 cm} Fig.6: \, Asymptotically $z= -1$ stable equilibrium sink for $\gamma=-\frac{1}{30}$\\
\hspace{1 cm}  with initial values which have been considered by $ x(0) = 0.1$ and $y(0) = 0.07$.\\
\end{tabular*}\\
\\
\begin{tabular*}{2.5 cm}{cc}
\includegraphics[scale=.4]{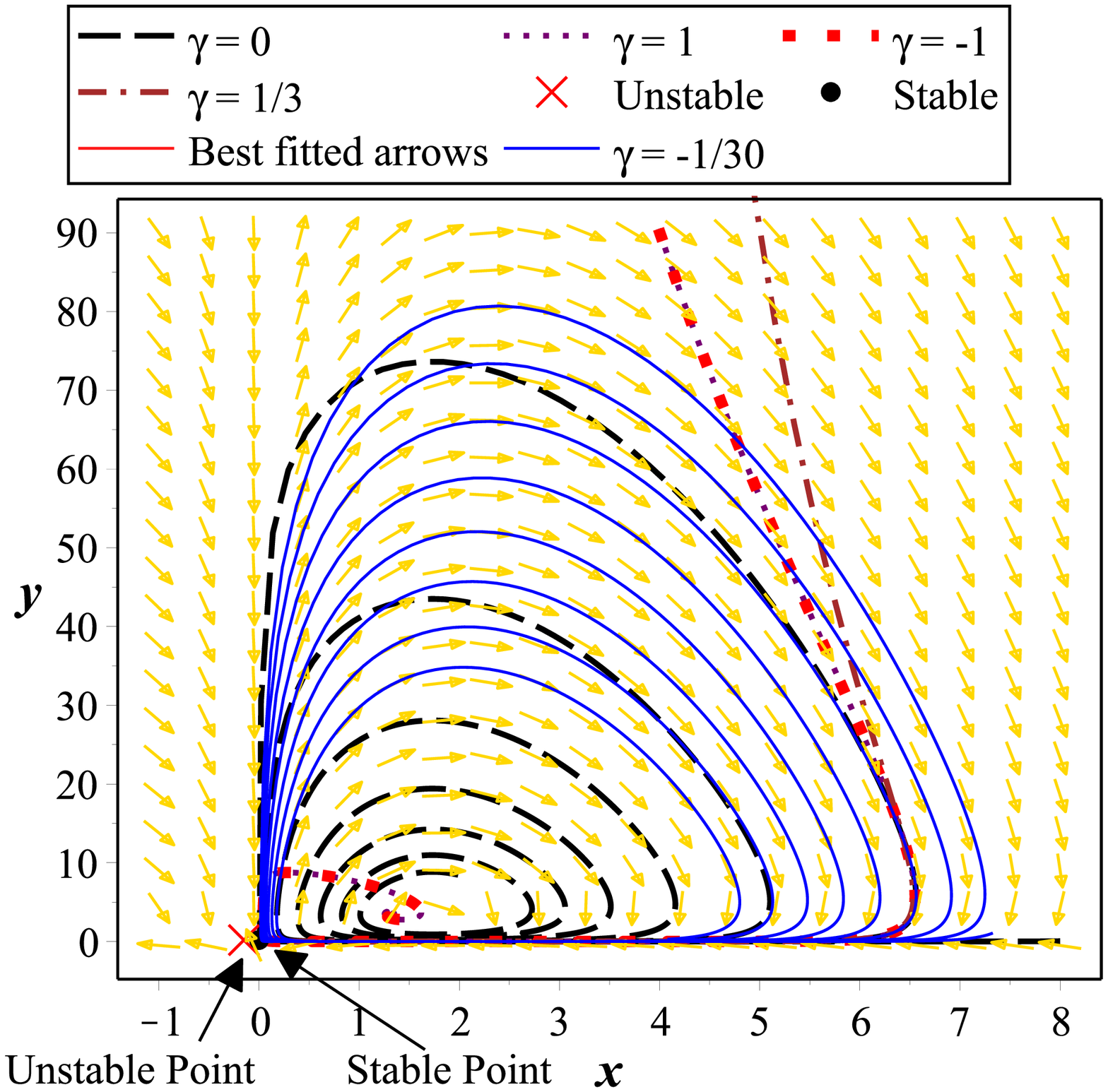}\hspace{0.5 cm}\includegraphics[scale=.4]{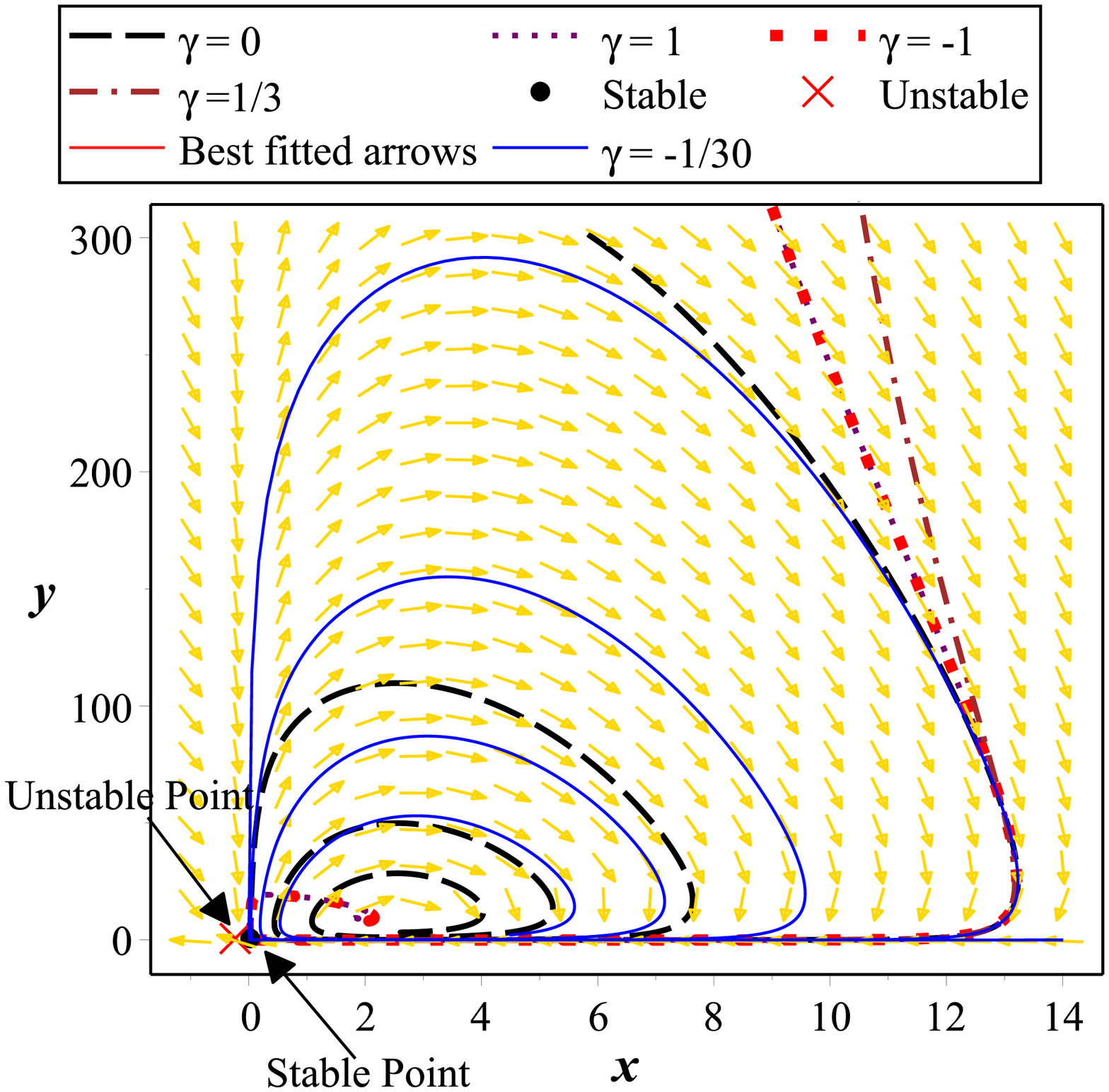}\\
\hspace{1 cm} Fig.7: \, Comparing of asymptotically $z= -1$ stable equilibrium sink for  different $\gamma$.\\
\hspace{1 cm}  The initial values have been considered by $ x(0) = 0.1$ and $y(0) = 0.07$. The left side \\
\hspace{1 cm} parameters are $\delta_1 = -10$ and $\delta_2 = -30$ while the right side they are $\delta_1 = -3$ and $\delta_2 = -8\sqrt{6}$.\\
\hspace{1 cm} Stable and unstable points have been shown for $\gamma=-1/30$\\
\end{tabular*}\\

\section{Cosmological Parameters}\label{sec:CP}
In order to the investigate the finial fate of the universe and which of different dark energy
models is now emergent, we should investigate the effective
EoS and deceleration parameters in terms of the dynamical variables in the universe and a sensitive and robust
diagnostic test for dark energy. In this way, the EoS and deceleration parameters are refined by:
\begin{eqnarray}
\omega_{eff}&\doteq&-1-\frac{2}{3}\frac{\dot{H}}{H^2}=\frac{\sqrt{6}}{3}\,k+1,\\
q&\doteq& -1-\frac{\dot{H}}{H^2}=\frac{\sqrt{6}}{2}\,k-1,
\end{eqnarray}
and we can rewrite the statefinder diagnostic parameters, $\{ r, s \}$, as the follow,
\begin{eqnarray}
r&\doteq& \frac{\ddot{H}}{H^3}-3\,q -2=-\frac{3}{2}\left(3+\frac {1}{x}\right)k^2-3\sqrt {6}\,k-3\left({\delta_1}^{2}\gamma y+{\delta_2}^{2}z\right)+1,
\\
s&\doteq& \frac{r-1}{3\,\left(q-\frac{1}{2}\right)}=\frac{-2}{\sqrt{6}\,k-3}\left(\sqrt{6}\,k+(\delta_1^2\gamma\,y+\delta_2^2z)+\frac{k^2}{2}\left(3+\frac{1}{x}\right)\right),
\end{eqnarray}
%p&\doteq& -\frac{H^2}{\dot{H}}=\frac{\sqrt{6}}{3\,k}
which introduced by Sahni et al.\citep{sahni2003statefinder} and Alam et al.\citep{alam2003exploring}. On the other hand, from \citep{feng2008statefinder} for statefinder and deceleration parameters in Ricci dark
energy (RDE) framework, we have
\begin{eqnarray}
r &=& 1-\left(\frac{1}{\alpha^2}\right)\frac{(2-\alpha)(2\alpha-1)\beta e^{-(4-\frac{2}{\alpha})N}}{\frac{2}{2-\alpha}\Omega_{m0}e^{-3N}+\beta e^{-(4-\frac{2}{\alpha})N}},\\
s &=& \frac{2}{3}\left(2-\frac{1}{\alpha}\right),\\
q &=& \frac{1}{2}\left(1-\left(\frac{1}{\alpha}\right)\frac{(2-\alpha)\beta e^{-(4-\frac{2}{\alpha})N}}{\frac{2}{2-\alpha}\Omega_{m0}e^{-3N}+\beta e^{-(4-\frac{2}{\alpha})N}}\right)\cdot
\end{eqnarray}
So with current observations data and our previous calculations, one can find the location of today's point is $r=1.422$, $s=-0.131$ and $q=-0.578$.

In Fig. 8, we show the evolution of EoS and deceleration parameters respect to $N=\ln(a)$. It's clear that their behaviours are very similar ($\propto k$) but when we compare two coloured lines, which their conditions have been chosen ordinary, with each other, in each plot separately, we can find the black line with a $\gamma\preceq 0$ has a better cross at the current value against the red line which has a bit distance from it. Moreover, both of lines show the deceleration parameter is negative now but it tends to approach to zero and its positive value in the future however its slope is decreasing for a certain period of time. Its mean is, our universe is in a smooth acceleration expansion now but in the future its acceleration approaches to zero and then it will be negative. In the additional, the red line  crosses $q=0$ for two times and finds a positive acceleration again whereas the black line has just one crossed point in the near future, a period with approximately constant negative acceleration and a period with ascending deceleration parameter which play us the role of a big crunch. So, our universe will collapse in on itself into a quasi-singularity
as gravity pulls matter down, in the future.\\     

\begin{tabular*}{2.5 cm}{cc}
\includegraphics[scale=.35]{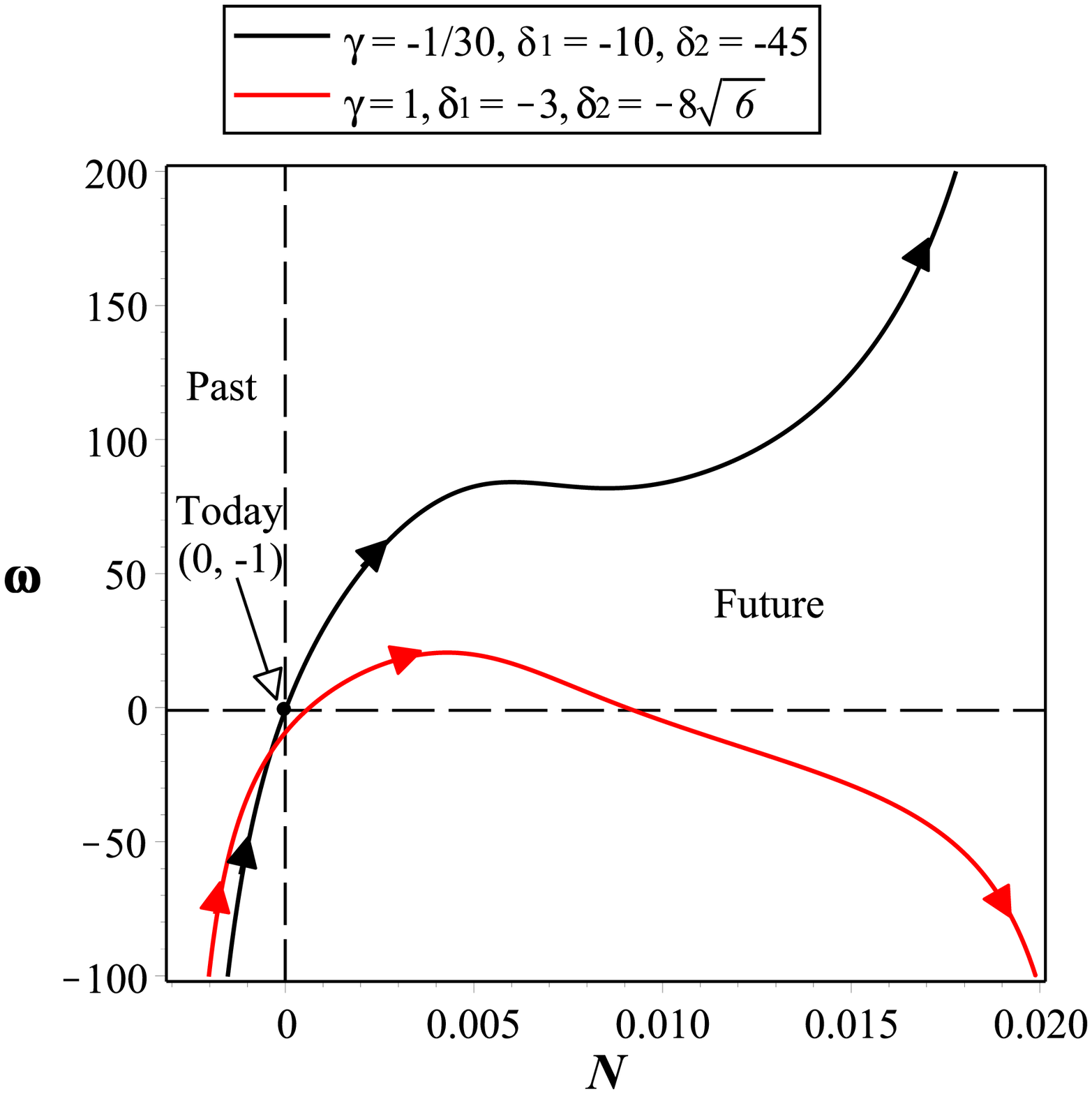}\hspace{1 cm}\includegraphics[scale=.35]{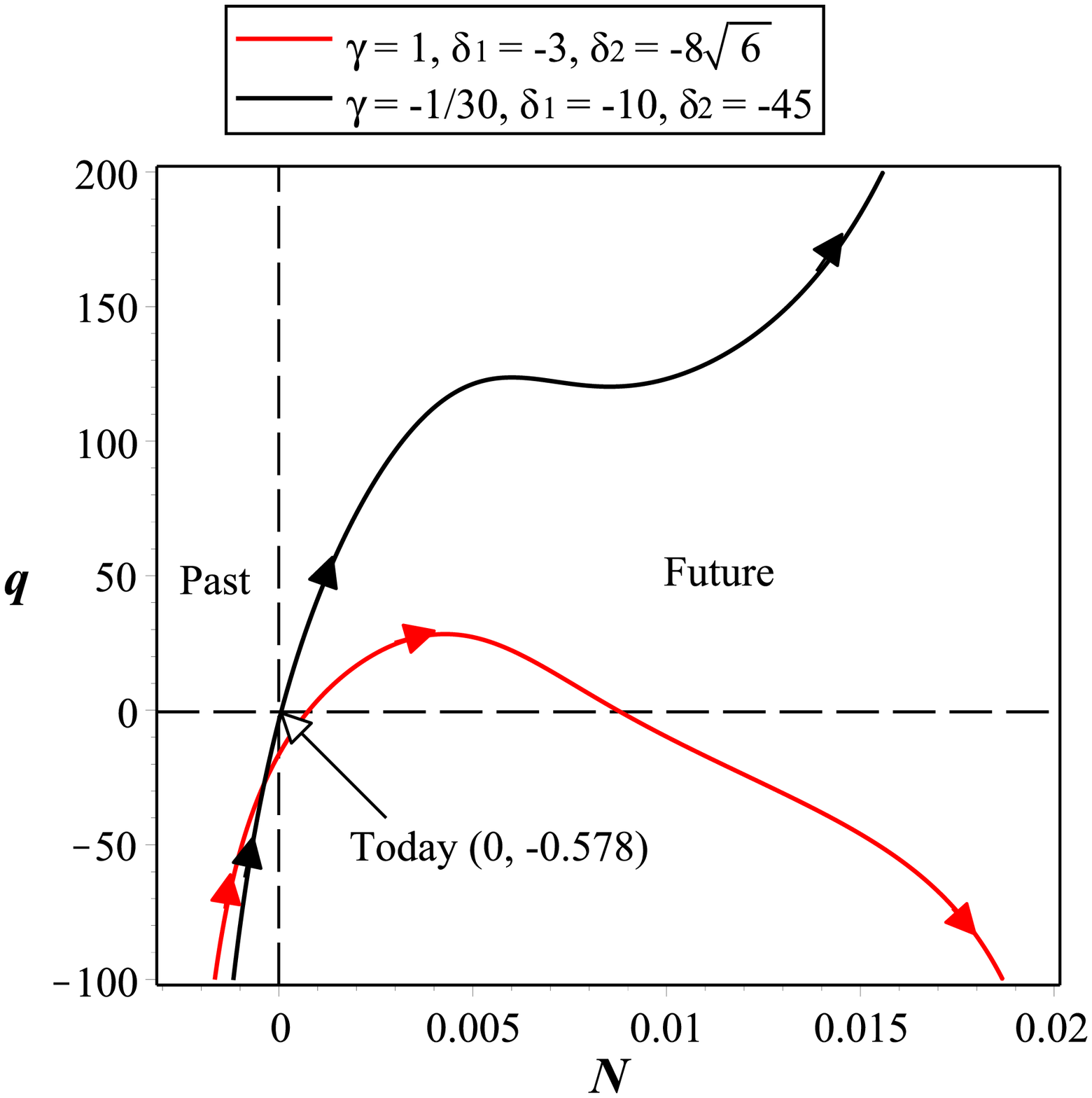}\hspace{4 cm}\\
\hspace{1 cm} Fig.8: \, Plot of the evolution of the EoS and deceleration parameters as a \\
\hspace{1 cm} function of $N=\ln(a)$, with initial values as, $x(0) = \sqrt{6}$, $y(0) = 9$, $z(0)=0.2$.\\
\end{tabular*}\\

In this way, The trajectory of lines indicate us, they cross $\omega=-1$ at a point near of the current time. Specially, the black one that crosses it exactly at the today's location. They run over from phantom phase, $\omega<-1$, to quintessence, $\omega>-1$ that emphasis the validation of our model. However, the red one will come back to its phantom phase again.   

On the other hand, the statefinder diagnostic can discriminate between various dark energy models effectively.
Different cosmological models involving dark energy exhibit qualitatively different
evolution trajectories in the $s - r$ plane. For example, the $\Lambda$CDM scenario corresponds to
the fixed point $s = 0, r = 1$ and the SCDM (Standard Cold Dark Matter)
scenario corresponds to the point $s = 1, r = 1$. In Fig. 9, we try to show the treatment of the statefinder parameters, $s$ and $r$, separately respect to $ N = \ln (a)$. The $s-N$ plot shows us the black line has relatively a simple behaviour from past to future. However, it has a very sharp and short turning in the present time but the red line will experience two deep singularity at the future. Moreover, in the $r-N$ plot we can see the simplest behaviour is for the black one again.\\ 

\begin{tabular*}{2.5 cm}{cc}
\includegraphics[scale=.35]{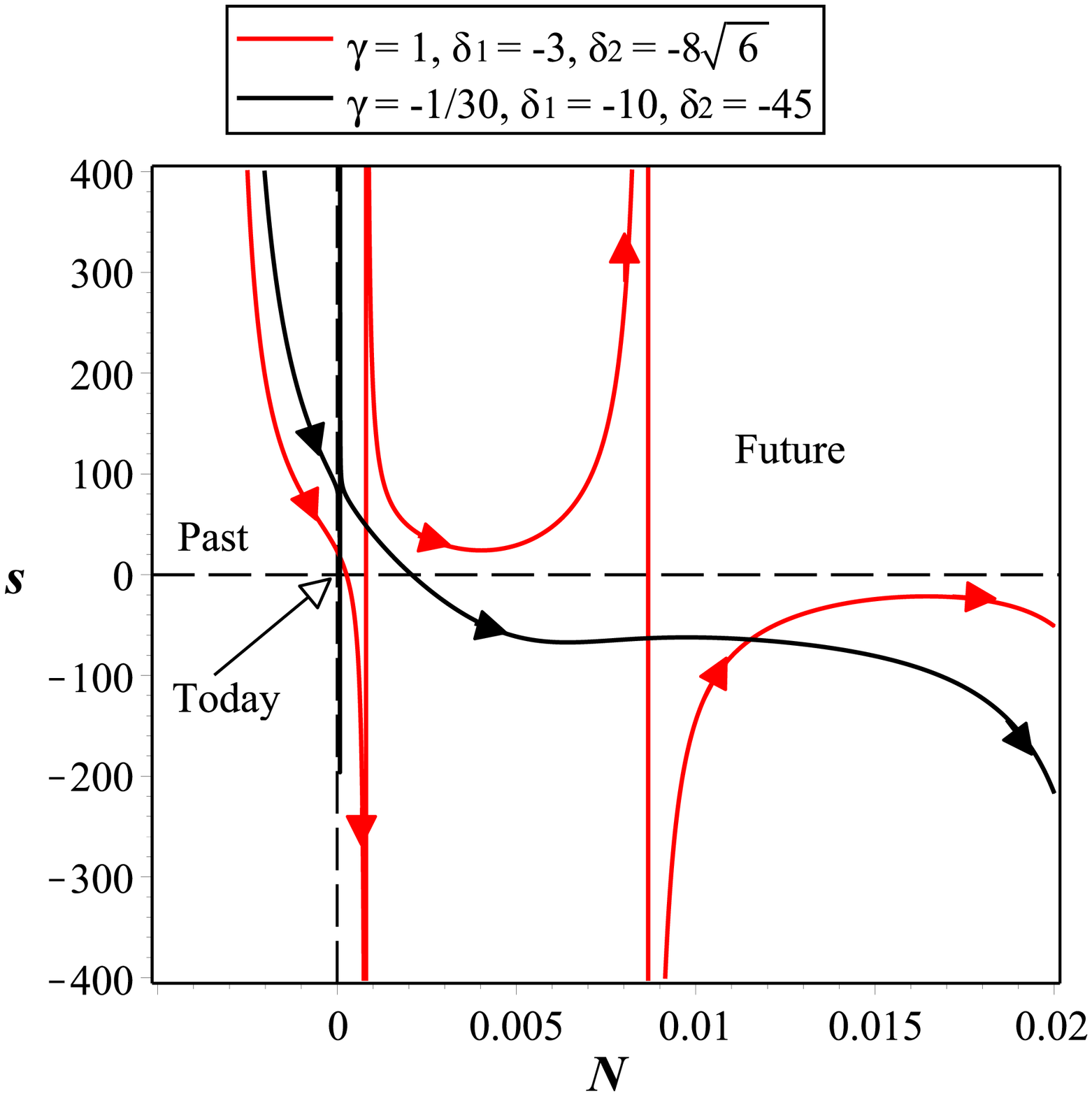}\hspace{1 cm}\includegraphics[scale=.35]{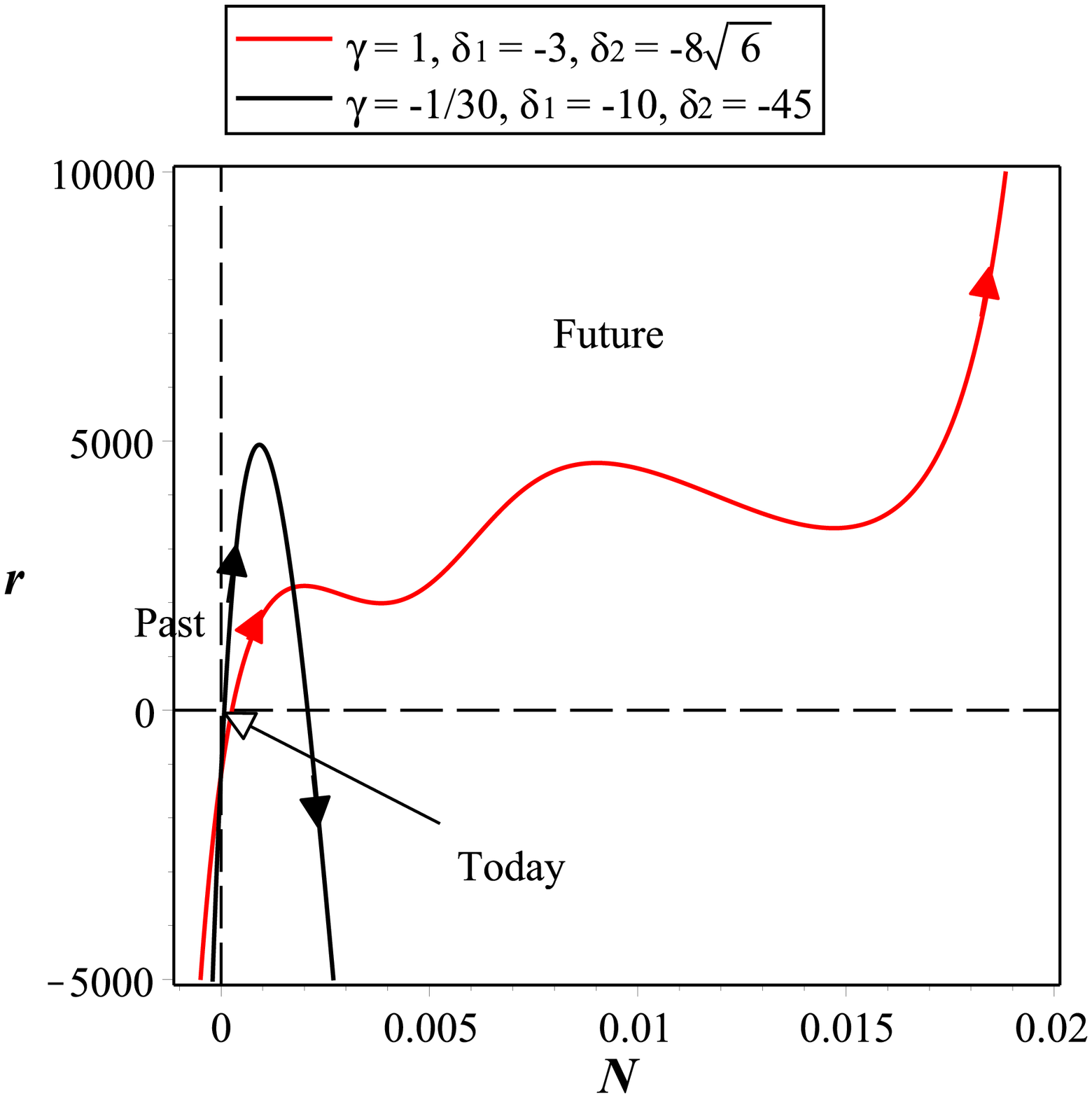}\hspace{4 cm}\\
\hspace{1 cm} Fig.9: \, Plot of the evolution of the statefinder parameters separately as a \\
\hspace{1 cm} function of $N=\ln(a)$, with initial values as, $x(0) = \sqrt{6}$, $y(0) = 9$, $z(0)=0.2$.\\
\end{tabular*}\\

In Fig. 10, we've plotted the $s - r $ plane of our model with two ordinary different conditions. For showing more detailed and finding the discrimination of different models we show it in two scale. The left one is closer and shows the black trajectory which has a $\gamma\preceq 0$, has closer distance to current time and SCDM against the red line which has a $\gamma=1$. However both of them cross on $\Lambda$CDM. Therefore, the non-distance from our model to
the $\Lambda$CDM scenario can be identified explicitly. The distinctive trajectories which various
dark energy scenarios follow in the $s - r$ plane demonstrate quite strikingly the contrasting
behaviour of dark energy models.\\ 
  
\begin{tabular*}{2.5 cm}{cc}
\includegraphics[scale=.35]{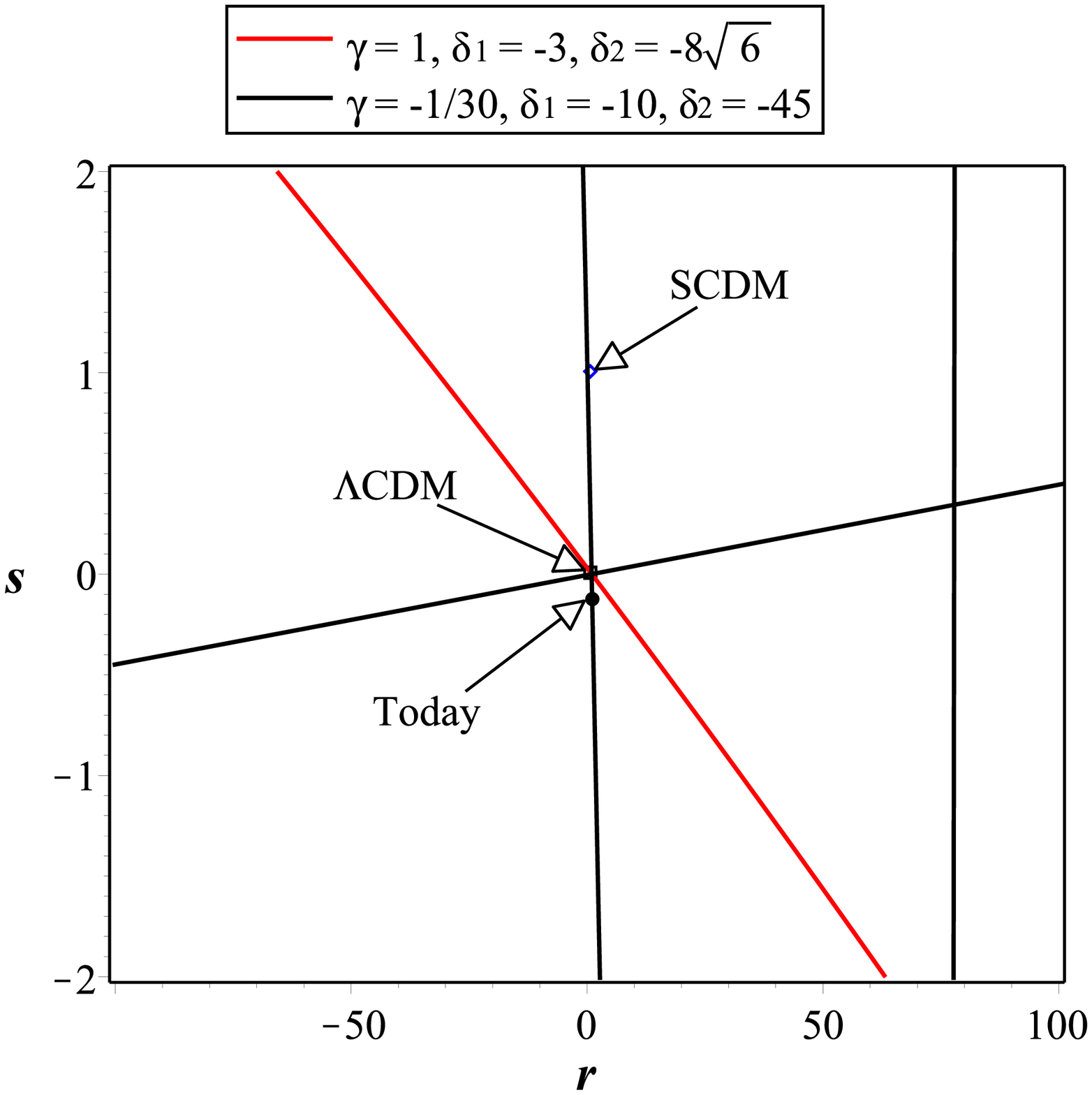}\hspace{1 cm}\includegraphics[scale=.35]{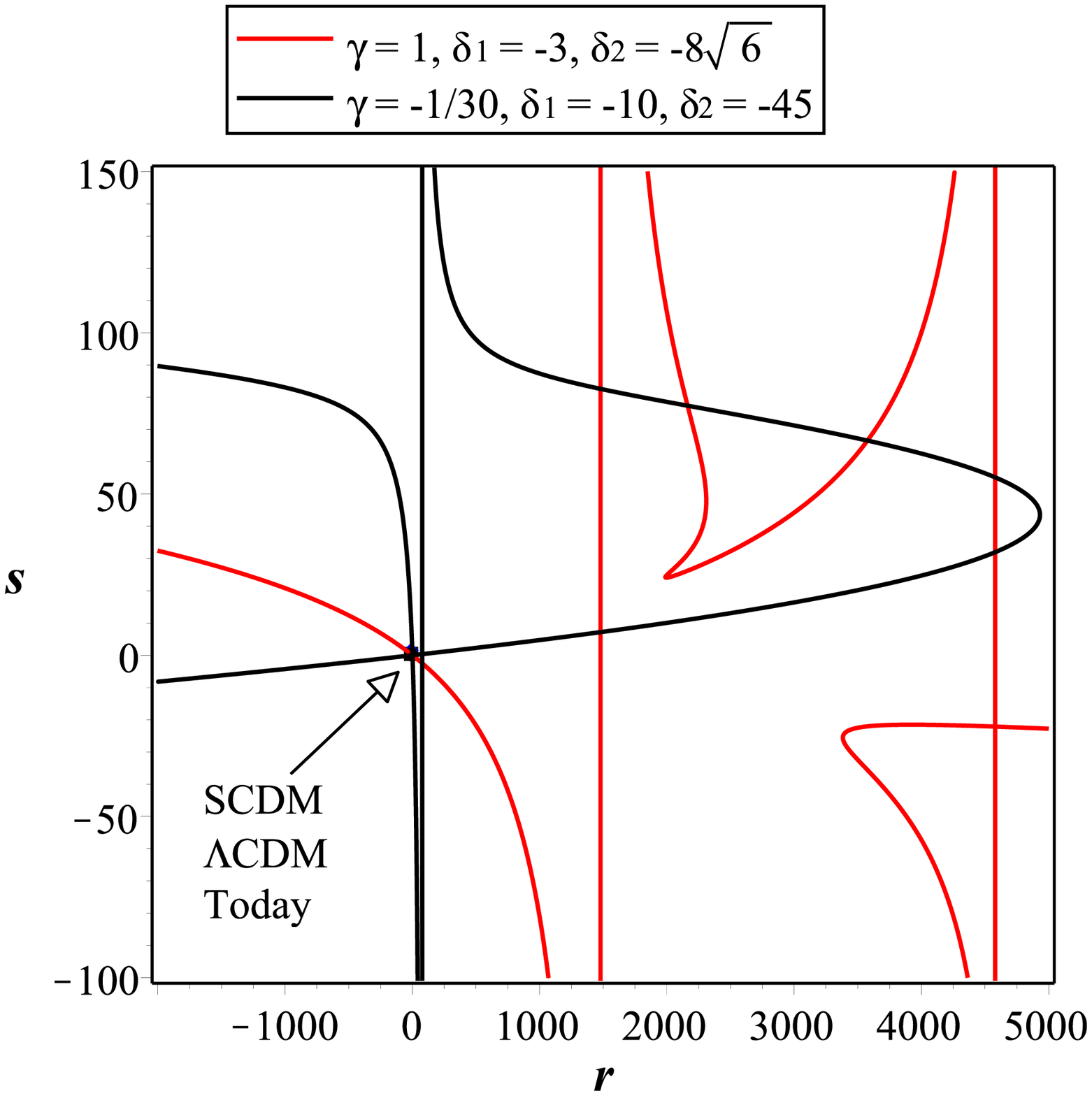}\hspace{4 cm}\\
\hspace{1 cm} Fig.10: \, Plot of the phase plane of $s$ parameter respect to $r$ in 2 scales.\\
\hspace{1 cm} The initial values are $x(0) = \sqrt{6}$, $y(0) = 9$, $z(0)=0.2$.\\
\end{tabular*}\\

As a complementary, Fig.11, shows another statefinder diagrams, the $s(q)$ and $r(q)$ evolutionary
trajectories. From left side of this figure, we clearly see that the black line is more closer to $\Lambda$CDM scenario ($s = 0 , q = -0.5$), current time ($s = -0.131 , q = -0.578$) and SCDM ($s = 1 , q = 0.5$), while the SCDM scenario is located just in the red line singularity. But the steady state cosmology
(SS) which explains the de-Sitter expansion, with $s = 1$ and $q = -1$, has more distance from our model in both of conditions. 

On the other hand, from right side of this figure, we could able to see that the black line is more closer to $\Lambda$CDM scenario ($r = 1 , q = -0.5$), current time ($r = 1.422 , q = -0.578$) and SCDM ($r = 1 , q = 0.5$) again and SS scenario ($r = 1 , q = -1$) has the most distance from our model against other models. In this figure, the red line has been separated from all scenarios. Therefore, the definition conditions of the red line will be failed for introducing a suitable method and the winner of our match in the most of circumstances, will be stayed the definition conditions of the black line.\\

\begin{tabular*}{2.5 cm}{cc}
\includegraphics[scale=.35]{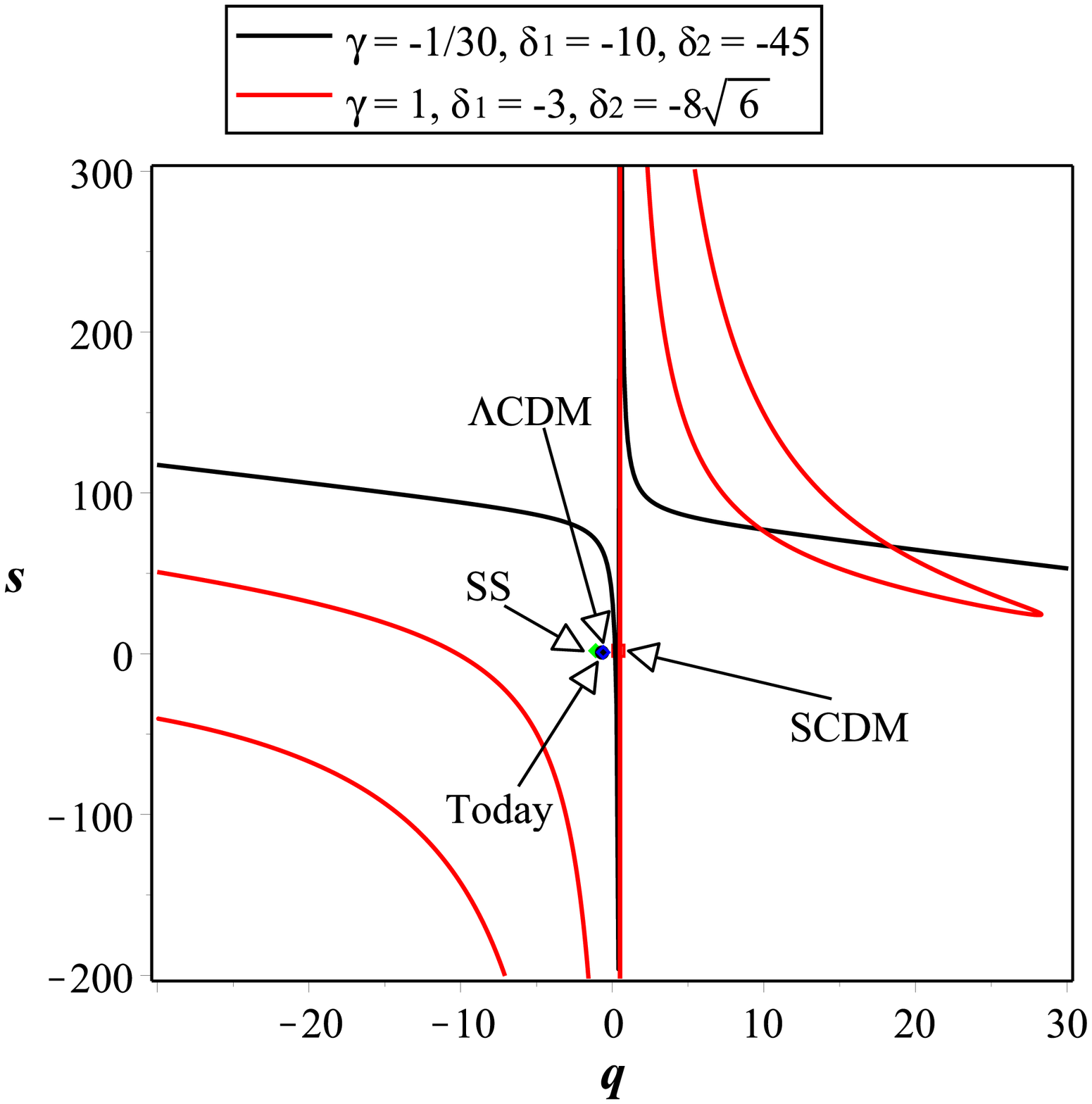}\hspace{1 cm}\includegraphics[scale=.35]{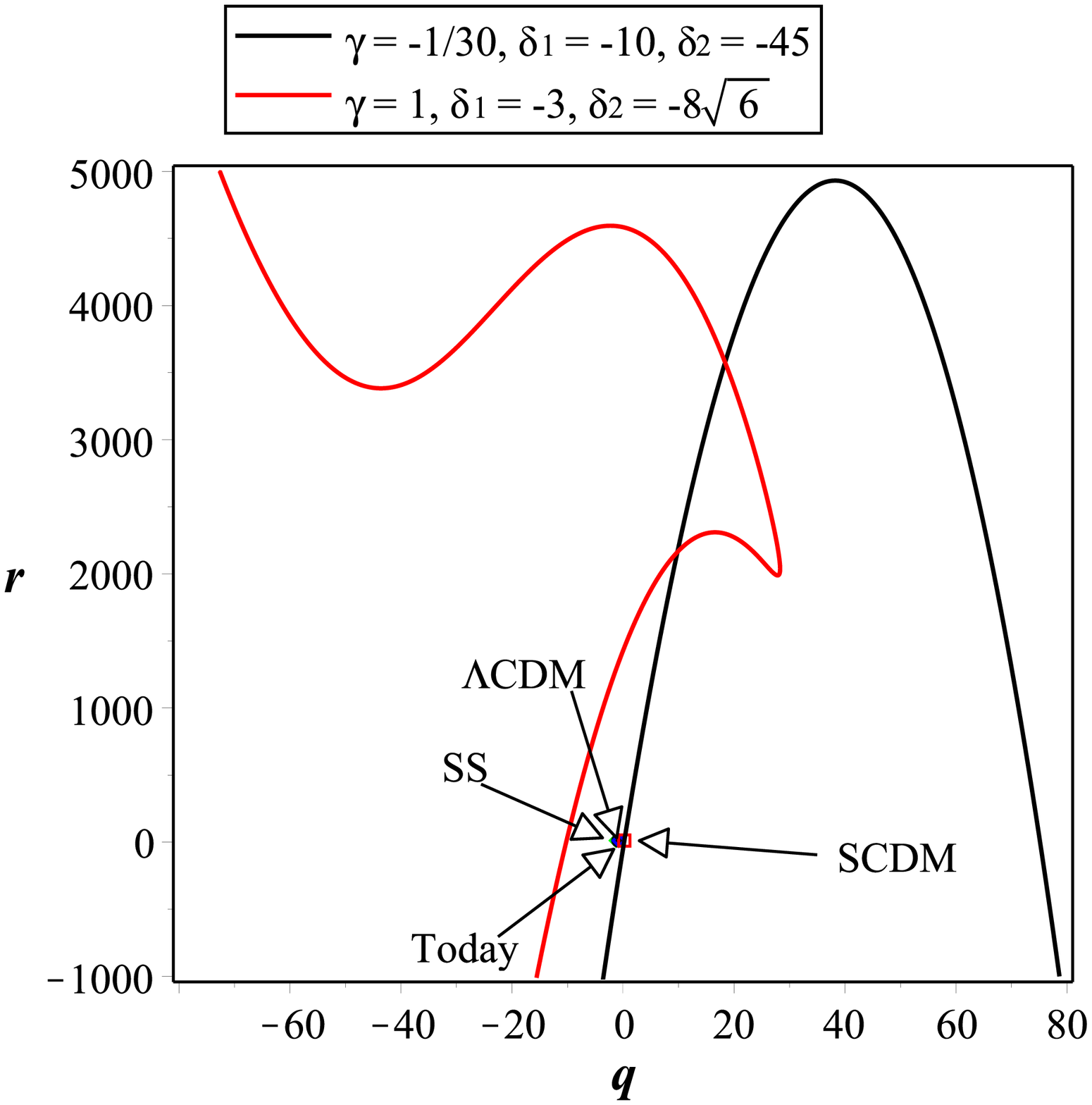}\hspace{4 cm}\\
\hspace{1 cm} Fig.11: \, Plot of the phase plane of $s$ and $r$ parameters respect to $q$.\\
\hspace{1 cm} The initial values are $x(0) = \sqrt{6}$, $y(0) = 9$, $z(0)=0.2$.\\
\end{tabular*}\\

Finally for having the analytically answers for our cosmological parameters with an arbitrary $\gamma \preceq 0$,  we rewrite them as table II,

\begin{table}[ht]
\caption{Cosmological parameters for $\gamma=-\frac{1}{30}$} % title of Table
\centering % used for centering table
\begin{tabular}{c c c c c c} % centered columns (4 columns)
\hline\hline %inserts double horizontal lines
$\textbf{Points}$ & $\textbf{q}$ & $\omega_{eff}$ & $\textbf{r}$ & $\textbf{s}$\,\,\,\,\,\,\,\, & $\textbf{scalar factor}$\\ [0.5ex] % inserts table
%heading
\hline % inserts single horizontal line
$\textbf{P}_\textbf{1}\textbf{:}$ & $\frac{-41-26\sqrt{6}}{5+5\sqrt{6}}$\,\,\, & $\frac{-19-9\sqrt{6}}{5+5\sqrt{6}}$&\,\,\,\,\,\,\,\, $r_1$\,\,\,\,\,\,\,\,&
$s_1$\,\,\,\,\,\,\,\,
& $a\propto t^{\frac{6-\sqrt{6}}{18}}$\\ [1ex]% inserting body of the table
$\textbf{P}_\textbf{2}\textbf{:}$ & $-1+\frac{\sqrt{6}}{10\sqrt{6}-10}$\,\,\, & $1+\frac{\sqrt{6}}{15\sqrt{6}-15}$\,\,\,\,\,\, & $r_2$
& $s_2$\,\,\,\,\,\,\,\,\,
&$a\propto t^{\frac{5(6-\sqrt{6})}{3}}$\\ [1ex] % [1ex] adds vertical space
\hline %inserts single line
\end{tabular}
\label{table:nonlin} % is used to refer this table in the text
\end{table}
where
\begin{eqnarray}
r_1 &=& \frac{1}{25}\,{\frac { \left( -1260\,\delta_2+1341 \right) \sqrt {6}-3285\,
\delta_2+2581}{ \left( 1+\sqrt {6} \right) ^{3}}}
,\nonumber\\
s_1 &=&-\frac{2}{5}{\frac { \left( 2\,\sqrt {6}+7 \right)  \left( 16\,\sqrt {6}+6-
10\,\delta_2\,\sqrt {6}-35\,\delta_2 \right) }{ \left( 29+19\,\sqrt {6}
 \right)  \left( 1+\sqrt {6} \right) ^{2}}},
\nonumber\\
r_2&=& {\frac {1}{50}}\,{\frac { \left( 420\,\delta_1-2833 \right) \sqrt {6}-
1095\,\delta_1+6953}{ \left( 2\,\sqrt {6}-7 \right)  \left( \sqrt {6}-1
 \right) ^{3}}}
,\nonumber\\
s_2&=&\frac{1}{5}\,{\frac {1699-739\,\sqrt {6}+365\,\delta_1-140\,\delta_1\,\sqrt {6}
}{ \left( 2\,\sqrt {6}-7 \right)  \left( 14\,\sqrt {6}-15 \right)
 \left( \sqrt {6}-1 \right) ^{2}}}
\cdot
\end{eqnarray}

It's clear that, we can find the critical points for each conditions as black line's easily.

\section{Summary and Conclusion}\label{sec:SumCon}
In this paper, the evolution of the gravitational fields was considered by the $f(R,\phi)$ modified gravity model which
coupled by the first gravitational theory where it
was invariant under the scale transformation and was presented by
Weyl. A formulation of gravity, as a simple modified model was
characterized by one scalar field $\phi$. The analytic study of
the solution shows, that the universe may undergo a transition
from phantom to quintessence phase which is also supported by
numerical analysis. In the analytic studying of the dynamics of
the EoS parameter, we obtain the constraints that one has to
impose on the scalar field and their first and second derivatives
in order to have phantom crossing. In numerical approach, the EoS
parameter crosses $\omega=-1$ for $t = 0$ or $\ln(a)= 0$. We
investigated about a bouncing non-singular cosmology, with an
initial contracting phase which lasts until to a non-vanishing
minimal radius is reached and then transits into an expanding
phase which provides a possible solution to the singularity
problem of Standard Big Bang cosmology, a problem which is not
cured by scalar field driven inflationary models. The evolution of
EoS parameter, Hubble parameter and scale factor numerically are
obtained. The violations of the null energy condition required to
get a bounce are obtained for the model allowing a transition of
the EoS parameter through the cosmological constant boundary. 
In the analytic discussion of the phantom
crossing behavior of the EoS parameter, we also have to constrain
the scalar field and their first and second derivatives. Besides,
we have also additional constraints on Hubble parameter and its
first and second derivatives.

Furthermore, we consider the structure of the dynamical system via phase plane analysis,
by introducing some dimensionless variables. By using some of constraints we removed 2 of 5 dimensionless variables. Then we found the critical points and imposed them a linear perturbation. By finding their eigenvalues, we discussed about the necessary conditions of stability both analytically and numerically. We plotted asymptotically $z= -1$ stable equilibrium sink for  different $\gamma$. The results indicated us the suitable conditions for a successful stability will be happened when we have $ \gamma\preceq 0$.

As an investigation to finding the finial fate of the universe and which of different dark energy
models is now emergent, we investigated the effective
EoS and deceleration parameters in terms of the dynamical variables in the universe. In this way, the EoS and deceleration parameters was considered for two ordinary conditions which one of them has a $ \gamma\preceq 0$ and other one didn't have. By using the equations which obtained from RDE framework and some observation data, we obtained the today's location of $s$, $r$ and $q$ parameters. We found that in the most of cases this condition gave us better responses because it has the most corresponding with today's values, specially $\Lambda$CDM and SCDM models. Besides, the statefinder diagnostic trajectories were plotted numerically for different situations. We discussed a lot why the first conditions is our winner. By a series of analysis, it was concluded, we must have minimally one crunch in the future and this reinforces the probability of being correct the hypothesis of big bounce and big crunch or cycling universe more.

\bibliographystyle{unsrt}
\bibliography{FMRef}
\end{document}